\documentclass[graybox]{svmult}

\usepackage{mathptmx}       
\usepackage{helvet}         
\usepackage{courier}        
\usepackage{makeidx}         
\usepackage{graphicx}        
\usepackage{multicol}        
\usepackage[bottom]{footmisc}
\usepackage{amsmath} 
\usepackage{mathtools} 
\usepackage{xcolor}
\usepackage{comment}
\usepackage{tikz-cd}
\usepackage{amsfonts}
\usepackage{kbordermatrix}
\usepackage{caption}

\newcommand{\p}{\mathbb{P}}

\makeindex             

\begin{document}

\title*{Classifying sleep states using persistent homology and Markov chain: a Pilot Study}
\author{Sarah Tymochko, Kritika Singhal, and Giseon Heo}

\institute{
Sarah Tymochko \at Dept. of Computational Mathematics, Science and Engineering, Michigan State University, East Lansing, MI, \email{tymochko@egr.msu.edu} 
\and
Kritika Singhal \at Dept. of Mathematics, The Ohio State University, Columbus, OH, \email{k2singhal@gmail.com}
\and 
Giseon Heo (corresponding author) 
\at  School of Dentistry; Department of Mathematical and Statistical Sciences, University of Alberta, Edmonton AB T6G 1C9, Canada. \email{gheo@ulberta.ca}
}

%

%
%
\maketitle
\abstract{Obstructive sleep Apnea (OSA) is a form of sleep disordered breathing characterized by frequent episodes of upper airway collapse during sleep. 
Pediatric OSA occurs in 1-5\% of children and can related to other serious health conditions such as high blood pressure, behavioral issues, or altered growth. 
OSA is often diagnosed by studying the patient's sleep cycle, the pattern with which they progress through various sleep states such as wakefulness, rapid eye-movement, and non-rapid eye-movement.
The sleep state data is obtained using an overnight polysomnography test that the patient undergoes at a hospital or sleep clinic, where a technician manually labels each 30 second time interval, also called an ``epoch,'' with the current sleep state. This process is laborious and prone to human error. 
We seek an automatic method of classifying the sleep state, as well as a method to analyze the sleep cycles.
This article is a pilot study in sleep state classification using two approaches: first, we'll use methods from the field of topological data analysis to classify the sleep state and second, we'll model sleep states as a Markov chain and visually analyze the sleep patterns.
In the future, we will continue to build on this work to improve our methods.}

\section{Introduction}\label{sec:intro}
Obstructive sleep apnea (OSA) is a chronic condition characterized by frequent episodes of upper airway collapse during sleep. 
Pediatric OSA is a serious health problem; even mild forms of untreated pediatric OSA may cause high blood pressure, changes to the heart, behavioral challenges, or alter the patients’ growth. 
The prevalence of childhood OSA is in the range of 1-5\%.
The gold standard for diagnosis of pediatric OSA is by overnight polysomnography~(PSG) in a hospital or sleep clinic. 
PSG provides multi-channel time series, such as EEG, ECG, EOG, EMG, airflow, and SpO2. 
In order to determine the sleep state, a sleep technician must manually assign the sleep state for each epoch (a 30 second interval) based on the PSG multi-channel time series.
This is a laborious process which is subject to human error.

Following this labeling process, a sleep specialist determines the severity of OSA by assigning a \textit{apnea hypopnea index} (\emph{ahi}) based on the PSG time series and sleep patterns of a patient.
The severity of OSA in children is categorised as \emph{none} ($\rm{ahi} <1$), \emph{mild} ($ 1 \leq \rm{ahi} \leq 5$), \emph{moderate} ( $ 5 < \rm{ahi} \leq 10$), or \emph{severe} ($\rm{ahi} >10$).


There are three main types of sleep, wakefulness, rapid eye-movement (REM), and non rapid eye-movement (NREM).
The NREM is further divided into three states: NREM1, NREM2, NREM3. REM sleep, brain waves are similar to waking. 
NERM1 is transition between waking and sleep, NREM 1-2 are refereed as Stage 1-2 (light sleep) and NERM3 is Stage 3 (deep sleep).
For the majority of people, sleep begins with a short period of NREM1 then go to deeper sleep, and returning to REM. The progression through the various sleep stages is called a sleep cycle. The sleep cycle depends on age: 45-50 min for $<$ 2 years, 50-60 min for 2-4 years, 60-80 min for 5-11 years, 80-100 min for 12-13 years and 90-120 min for  $> 13$ years. The proportions of  REM and NREM also depend on age, proportion NREM increases as get older. Roughly, 50-50\% for newborn, 25-75\% for 6-30 years old, and  20-80\% for older than 30 years. 

\begin{figure}
    \begin{center}
     \scalebox{0.21}{\includegraphics{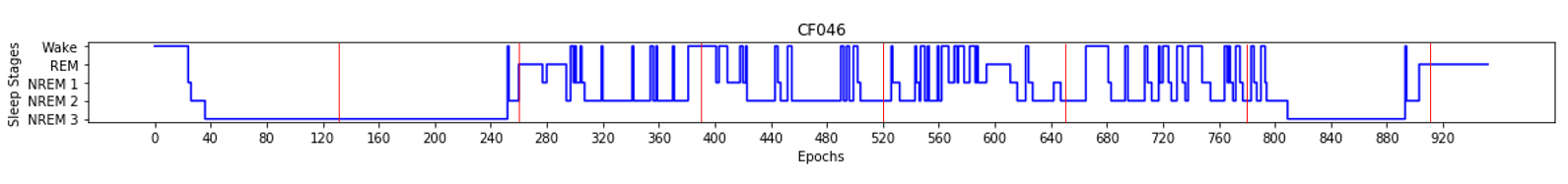}}
     \scalebox{0.21}{\includegraphics{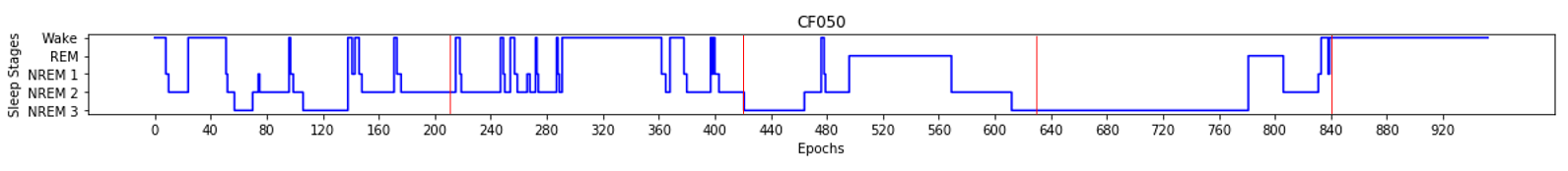}}
    \end{center}
    \caption{Hypnograms of patients CF046 and CF050 with sleep cycles 65 min~(130 epoch) and 105 min~(210 epoch) respectively. 
    }
    \label{fig:combinedhypnogram}
\end{figure}

Seventy eight children at risk of OSA were recruited and had taken polysomography~(PSG). This project was approved by the Health Research Ethics Board of the University of Alberta (Pro00057638).
As the objective of this article is exploring applications of persistent homology to multiple time series and understanding sleep stages and OSA severity, we will analyse a small subset of patients; eight patients and their PSG data. We  selected  two patients from each four OSA severity category. We present descriptive statistics of eight patients in the Table \ref{tab:sleepcycle}. 
\begin{table}[h]
    \centering
    \begin{tabular}{p{2.2cm}p{1cm}p{1cm}p{1cm}p{1cm}p{1cm}p{1cm}p{1cm}p{1cm}}
    \hline\hline 
	&	CF011	&	CF030	&	CF031	&	CF046	&	CF050	&	CF055	&	CF076	&	CF079	\\ \hline
Wake	&	24.47	&	7.81	&	5.31	&	13.58	&	25.67	&	15.18	&	17.00	&	24.53	\\
REM	&	14.16	&	19.22	&	17.93	&	11.22	&	10.48	&	17.24	&	13.70	&	8.41	\\
NREM1	&	2.37	&	1.30	&	2.47	&	8.76	&	3.64	&	4.41	&	5.10	&	7.32	\\
NREM2	&	40.27	&	41.58	&	35.50	&	32.43	&	32.73	&	41.92	&	41.60	&	40.26	\\
NREM3	&	18.72	&	30.09	&	38.79	&	34.01	&	27.49	&	21.25	&	22.60	&	19.49	\\
 \hline\hline
\emph{ahi} & 5.0 & 3.7 & 7.5 & 10.9 & 113.9& 40.2& 1.1 & 0.6 \\
Age (yr)    & 9.2& 4.1 & 4.1& 9.7&13.7&13.7&9.8&17 \\
Total sleep (hr)	&	9.13	&	10.88	&	9.11	&	9.51	&	7.80	&	8.51	&	8.33	&	8.43	\\
Duration of sleep cycle (min)	&	60-70	&	50-60	&	50-60	&	60-70	&	90-120	&	90-120	&	60-70	&	90-120	\\\hline
 \end{tabular}
 \caption{Percentage of each sleep stage, \emph{ahi}, age at the time of PSG was taken, total \# of sleep hours and duration of sleep cycle.}
 \label{tab:sleepcycle}
\end{table}
The sleep state recorded at each 30 second interval (epoch) over the entire sleep duration of a patient can be represented in a \emph{hypnogram}. 
The hypnograms of two patients, CF046 and CF050 are depicted in Figure \ref{fig:combinedhypnogram}. 

As labeling sleep state based on PSG data is currently done manually, it would be useful to come up with automated sleep scoring mechanism. 
This would optimize the effort and time of the technician, as well as remove the human error from the labeling process.
In this article, analyze time series variables from the PSG data using two separate approaches.
The first approach, discussed in section \ref{sec:PH}, using a persistent homology based method to predict the sleep state for 30 second interval of three time series.
For the second approach, in Section \ref{sec:markov}, we model sleep stage as 5-state Markov chain and visually inspect and sleep patterns of 8 patients.
The goal is to explore both the automatic prediction of sleep states, as well as observing relationships between OSA and sleep patterns.
We conclude our study with plan for future research.


\section{Sleep State Analysis using Persistent Homology}\label{sec:PH}

The utilization of tools from topological data analysis (TDA) for time series analysis has recently become a popular line of research. 
The combination of these fields has resulted in new methods for quantifying periodicity and distinguishing behavior in time series data.
PSG time series variables display different patterns for different sleep states.
One proposed method of detecting these differences is using persistent homology, a well studied tool for quantifying the underlying structure of data.
Persistent homology has been used to study time series data from many applications.
Existing applications include studying machining dynamics \cite{Khasawneh2015,Khasawneh2017,Khasawneh2018a}, gene expression \cite{Perea2015, berwald2014critical}, financial data \cite{Gidea}, and video data \cite{tralie2018quasi,tralie2016high}.
Additionally, \cite{Chung2019} used topological methods for time series analysis of sleep-wake states.
Our goal is to explore the use of persistent homology to classify sleep states in PSG data.
We also hope to see some variation in classification results that relate to the OSA severity of the patient.

\begin{figure}[h!]
\begin{tabular}{cc}
   \scalebox{0.15}{\includegraphics {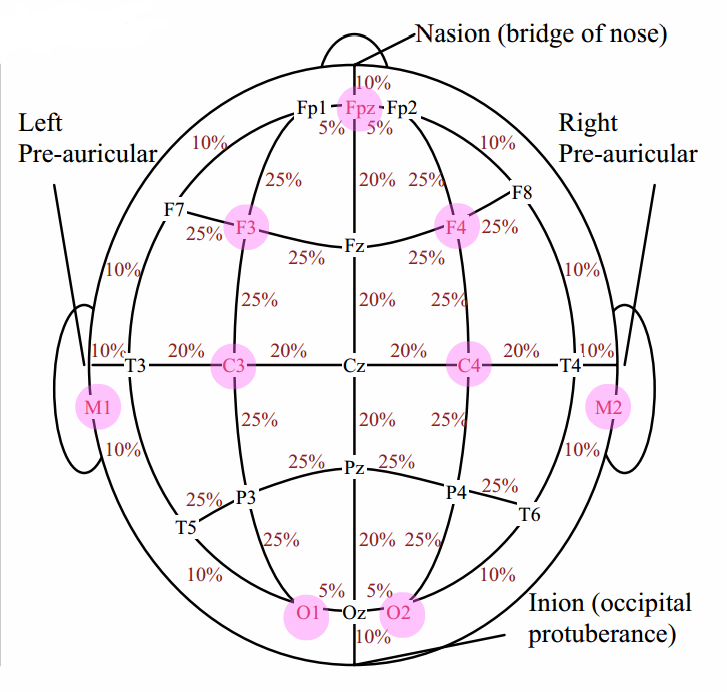}} & \scalebox{0.32}{\includegraphics{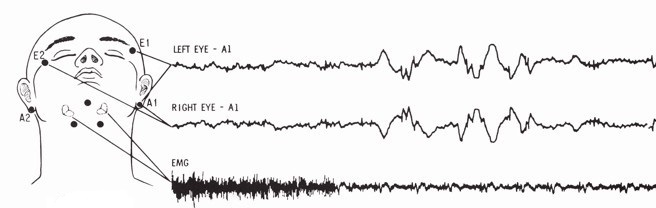}}
    \end{tabular}
    \caption{(Left) The International 10/20 System of Electrode Placement. Image from sleeptechstudy.wordpress.com.  (Right) EOG and EMG electrode placement. Image from \cite{R_K_Scoring_manual}}
    \label{fig:elec}
\end{figure}

As described in Sec.~\ref{sec:intro}, we have 8 patients with varying severity of OSA.
For each patient, we will consider three channels from the polysomnography data, specifically the central electrode  on the sclape (C3), left eye movement (LEOG), and right eye movement (REOG).
Each of these time series must be normalized by subtracting a reference electrode, M2 which placed on the mastoid. See Figure \ref{fig:elec}.

\subsection{Background} \label{ssec:PH_Background}

In this section, we'll present some necessary background information on persistent homology, time series analysis, and machine learning.
Section~\ref{sssec:PH} will cover the necessary background material, however for a more detailed overview, we refer the interested reader to \cite{Edelsbrunner2010, Hatcher, Munch2017,Oudot2015}.

\subsubsection{Persistent Homology} \label{sssec:PH}

A standard tool from algebraic topology is homology, which studies topological features such as connected components, loops and voids. 
Given a space $X$, homology computes a group, $H_p(X)$, for each dimension $p=0,1,2,\ldots$.
Each dimension contains information about the topological structure; specifically dimension 0 corresponds to connected components, dimension 1 corresponds to loops, and dimension 2 corresponds to voids. 
Here we'll focus on simplicial homology, but we must first define a few other concepts.

An $n$-simplex is defined as the convex hull of $n+1$ affinely independent points. 
For example, a 0-simplex is a vertex, a 1-simplex is an edge, and a 2-simplex is a triangle.
The face of an $n$-simplex, $\sigma$, is defined as the convex hull of a nonempty subset of the vertices in $\sigma$.
A simplicial complex $\mathcal{K}$ is a space built from simplices that satisfies two properties: first, the intersection of any two simplices in $\mathcal{K}$ must also be a simplex in $\mathcal{K}$ and second, all faces of a simplex in $\mathcal{K}$ must also be a simplex in $\mathcal{K}$.

For a simplicial complex $\mathcal{K}$, a $p$-chain, $c$, is a sum of $p$-simplices, $\sigma_i$ in $\mathcal{K}$, with some coefficients $a_i$, $c = \sum a_i \sigma_i$.
In this case, we're focusing on the simplified case of $a_i\in \mathbb{Z}_2$ as this is typically used for persistent homology. 
Here the collection of $p$-chains, or the chain group, denoted as $C_p(\mathcal{K})$, forms a vector space.
The boundary map is a linear transformation between chain groups $\partial_p:C_p\to C_{p-1}$ that maps a $p$-simplex to the sum of it's $(p-1)$-dimensional faces. 
The sequence of boundary maps between chain groups
\[
\cdots \xrightarrow{\partial_{p+1}} C_p \xrightarrow{\partial_{p}} C_{p-1} \xrightarrow{\partial_{p-1}} C_{p-2} \xrightarrow{\partial_{p-2}} \cdots
\]
Within chain groups, we define a $p$-cycle $c\in C_p$ with empty boundary, $\partial_p(c)=0$.
Thus, the set of $p$-cycles is the kernel of the boundary map, $\ker(\partial_p)$. A $p$-boundary is a $p$-chain $c_p\in C_p$, that is the boundary of a $p+1$-chain, $c_{p+1}\in C_{p+1}$, $c_p = \partial_{p+1} (c_{p+1})$.
The set of $p$-cycles is the image of the boundary map $\text{im}(\partial_p)$.
Then the $p$-th homology group is defined as 
\[
H_p(\mathcal{K}) = \ker(\partial_p)/\text{im}(\partial_{p+1}).
\]

Persistent homology is a method of studying the homology of a space across different scales. 
In this case, we will use the Vietoris-Rips complex to create a simplicial complex out of the point cloud. 
The Vietoris-Rips complex is defined for a point cloud, $X$ and a distance $r$, where for every finite set of $k$ vertices with maximum pairwise distance at most $r$, the $(k-1)$-simplex formed by those vertices is added to the complex.
Then taking a range of distance values, $\{r_i\}$ we get a set of simplicial complexes $\{X_{r_i}\}$ where if $r_i \leq r_j$, then $X_{r_i}\subseteq X_{r_j}$. 
Thus, taking an increasing sequence of distance values, $0 \leq r_0 \leq r_1 \leq \cdots \leq r_n$ results in a nested sequence of simplicial complexes,
\begin{equation}
X_{r_0} \subseteq X_{r_1} \subseteq \cdots \subseteq X_{r_n}     
\end{equation}
called a filtration.
The inclusions induce linear maps between the homology groups of each simplicial complex,
\begin{equation}
H_p(X_{r_0}) \to H_p(X_{r_1}) \to \cdots \to H_p(X_{r_n}).  
\end{equation}
These maps allow us to track how the homology changes through the filtration. 

A $p$-dimensional feature is ``born'' at the distance value corresponding to the first time in the sequence of homology groups that we see that feature appear.
More formally, $\gamma$ is born at $r_i$ if $\gamma \in H_p(X_{r_i})$ but $\gamma \not\in H_p(X_{r_{i-1}})$.
We say that a feature ``dies'' if it merges with an older feature.
Specifically, a feature $\gamma$ dies at $r_j$ if it merges with a feature that has earlier birth time between $X_{r_{j-1}}$ and $X_{j}$.
A persistence diagram is a method of recording this information, where a feature that is born at $r_i$ and dies at $r_j$ is represented by the point $(r_i,r_j)$. 

\begin{figure}[t]
    \centering
    \includegraphics[width=\textwidth]{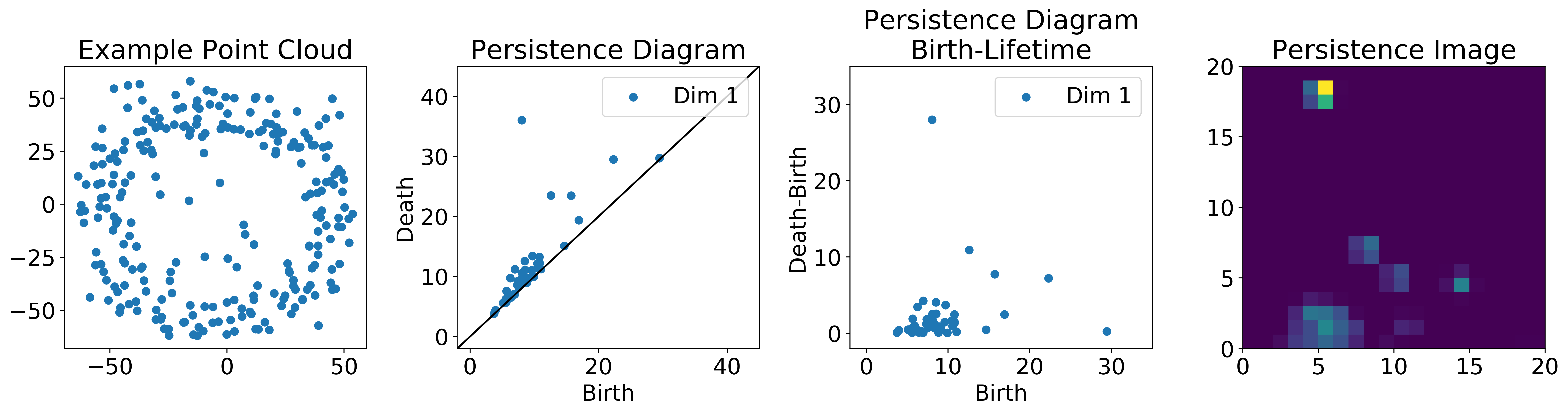}
    \caption{Example of a point cloud with corresponding persistence diagram and persistence image. The persistence diagram must first be transformed into birth-lifetime coordinates, meaning a point $(b,d)$ in the original diagram, is now plotted as $(b,d-b)$ in the birth-lifetime diagram. The persistence images method converts the birth-lifetime diagram into an image, represented as a $20\times 20$ matrix. That matrix can be flattened into a $1\times 400$ vector that can then be used for machine learning.}
    \label{fig:pd_pi}
\end{figure}

Persistence diagrams provide concise and robust summaries of the topological features on various scales. 
However, these diagrams are not well suited for machine learning tasks. 
There are many methods, often called featurization methods, of converting the information in a persistence diagram into a vector.
Once the diagram has been converted to a vector, it can be used in standard machine learning frameworks.  
In this paper, we'll use one featurization method called persistence images \cite{Adams2017}.
We will not go over the details of the method here, however an example can be seen in Fig.~\ref{fig:pd_pi}.

For implementation, we use scikit-tda python package \cite{scikittda2019} to compute persistent homology with ripser \cite{ctralie2018ripser} and to transform the persistence diagrams into persistence images with the Persim library.
For both of these computations, we use the default parameters.

\subsubsection{Time Series Analysis}

\begin{figure}[t]
    \centering
    \includegraphics[width=0.9\textwidth]{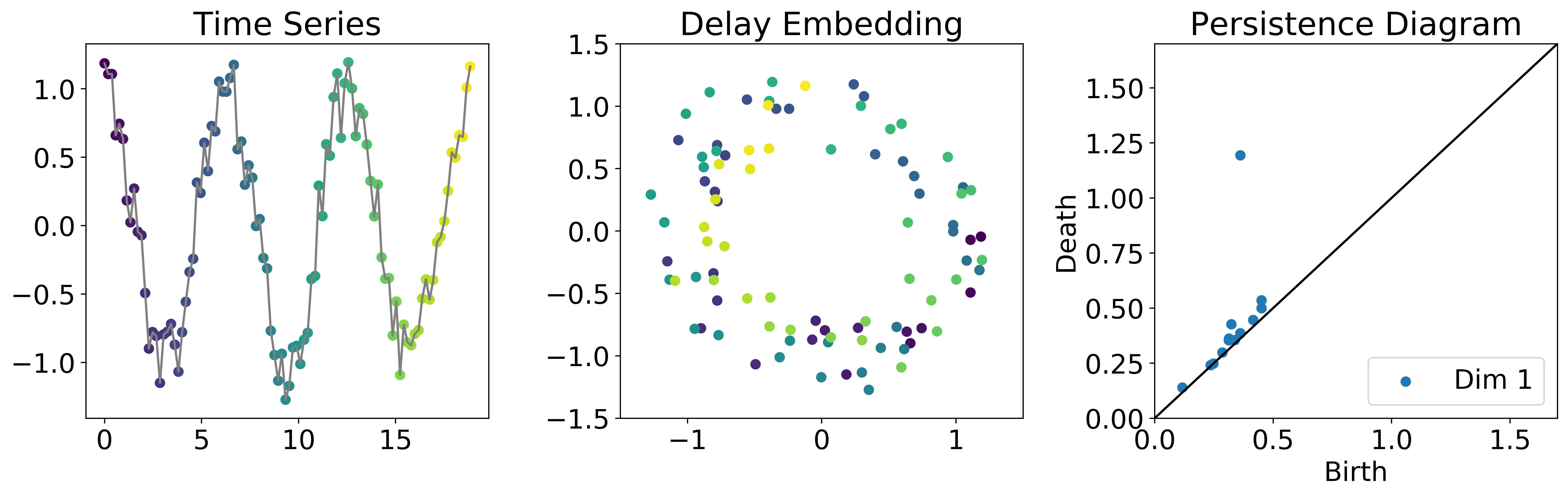}
    \caption{Example time series (left), delay embedding with $d=2$ and $\tau=9$ (middle), and corresponding persistence diagram (right). }
    \label{fig:Delay_Ex}
\end{figure}

Now that we have established a framework for persistent homology, we need to convert time series data into a form that is amenable to this type of analysis.
There are several existing methods to convert a time series into a point cloud.
One popular method, called a delay embedding, leverages Takens' theorem \cite{Takens1981}.
Given a time series, $X(t)$, we select two parameters, a dimension $d \in \mathbb{Z}_{>0}$ and a delay, $\tau>0$. 
The delay embedding is then defined as
 \[
 \phi_{\tau,d}: X(t) \mapsto (X(t), X(t+\tau), X(t+2\tau), \ldots, X(t+(d-1)\tau)
 \]
Takens' theorem proves that with the right parameters, this is in fact an embedding in the true mathematical sense as it preserves the underlying structure of the manifold.
This embedding is sensitive to the choices of $d$ and $\tau$. 
To choose these parameters automatically, we use a method based on permutation entropy, as presented in \cite{myers2019automatic}.
Once the time series has been embedded as a point cloud with this method, standard persistent homology can be applied. 
Figure~\ref{fig:Delay_Ex} has an example of the delay embedding method along with the corresponding persistence diagram.

\subsection{Results} \label{ssec:PH_Results}
 
 \begin{table}[t]
    \centering
    \begin{tabular}{|c|c|}
        \hline
        \textbf{\# of Classes} & \textbf{Labels}  \\
        \hline
        2 & Wake, Sleep \\
        3 & Wake, REM, NREM (1, 2\&3) \\
        4 & Wake, REM, Light Sleep (NREM 1\&2), Deep Sleep (NREM 3) \\
        5 & Wake, REM, NREM 1, NREM 2, NREM 3 \\
        \hline
    \end{tabular}
    \caption{Possible labels for 2, 3, 4 and 5 class classifications.}
    \label{tab:classes}
\end{table}

\begin{figure}
    \centering
    \includegraphics[width=0.9\textwidth]{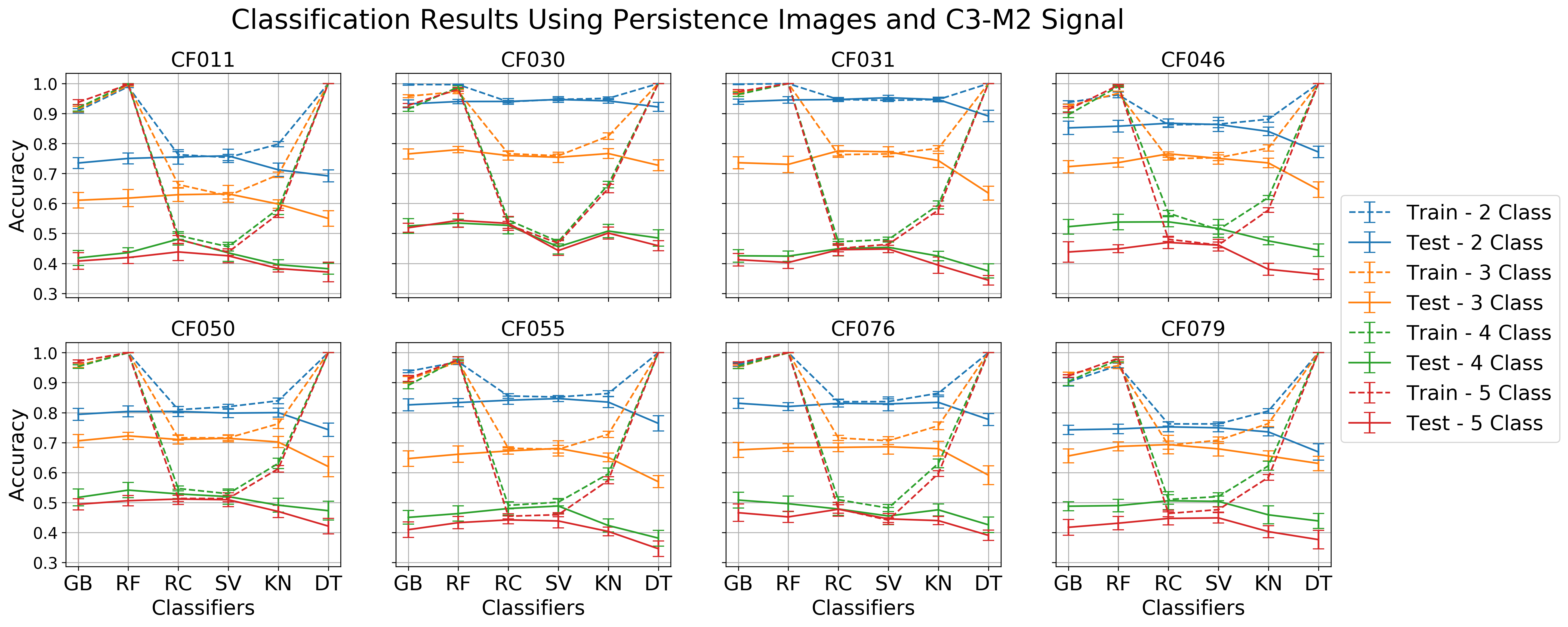}
    \includegraphics[width=0.9\textwidth]{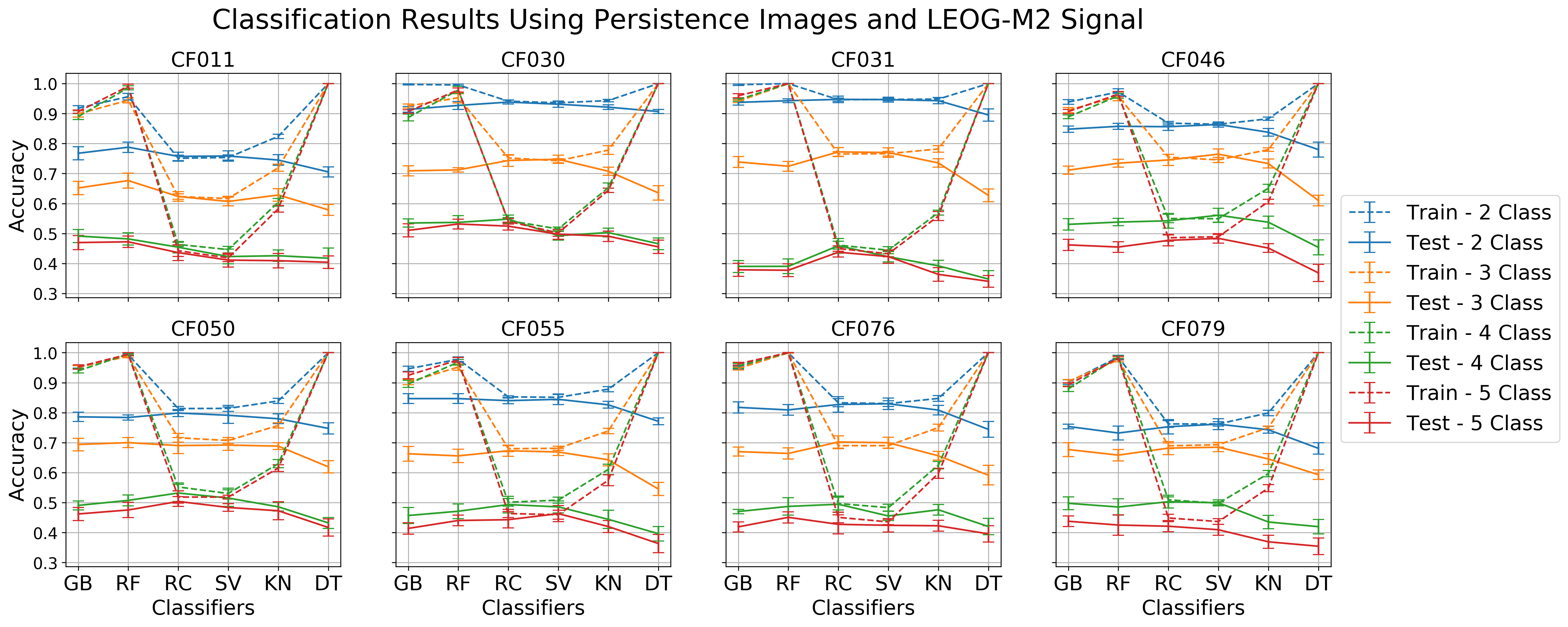}
    \includegraphics[width=0.9\textwidth]{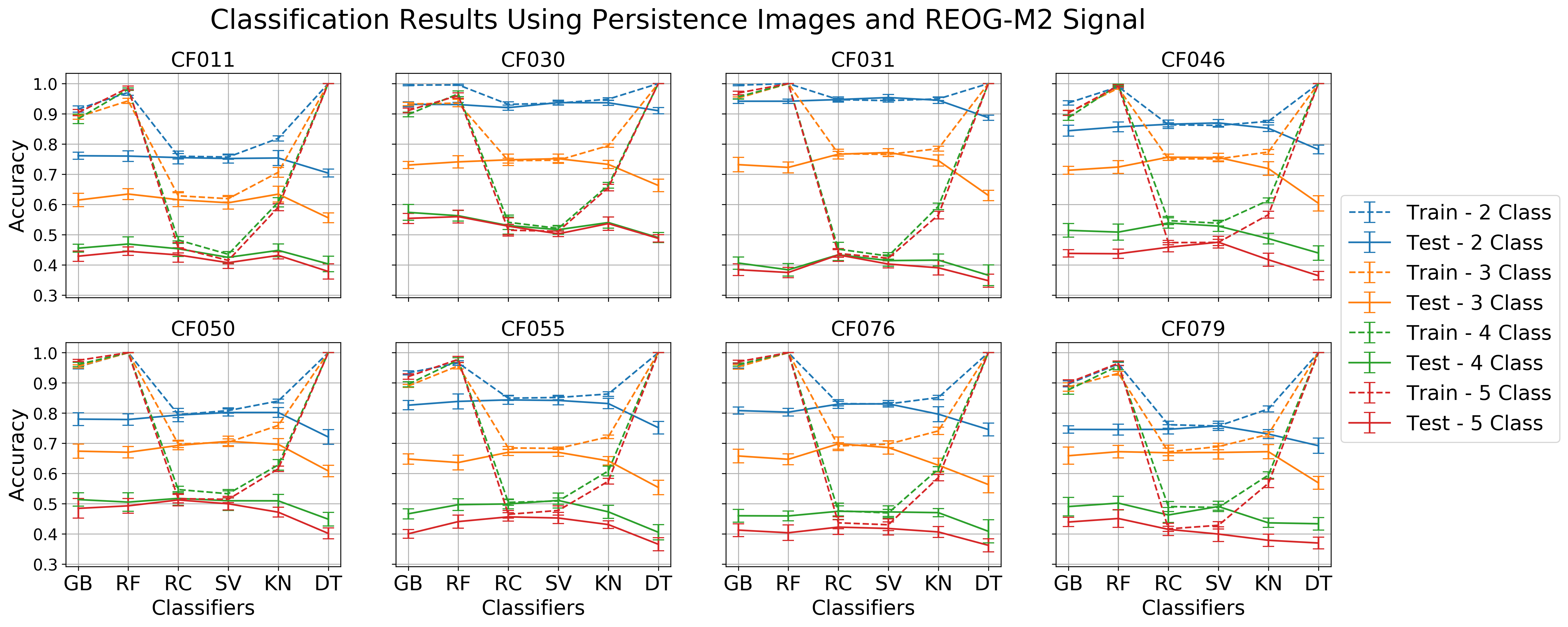}
    \caption{Average classification accuracies using persistence images for 2, 3, 4 and 5 classes for each patient. Error bars are the standard deviation. Classifiers used are gradient boosting (GB), random forest (RF), ridge classification (RC), support vector classifier (SV), K-neighbors classifier (KN) and decision tree (DT).}
    \label{fig:PI}
\end{figure}

\begin{figure}
    \centering
    \includegraphics[width=0.9\textwidth]{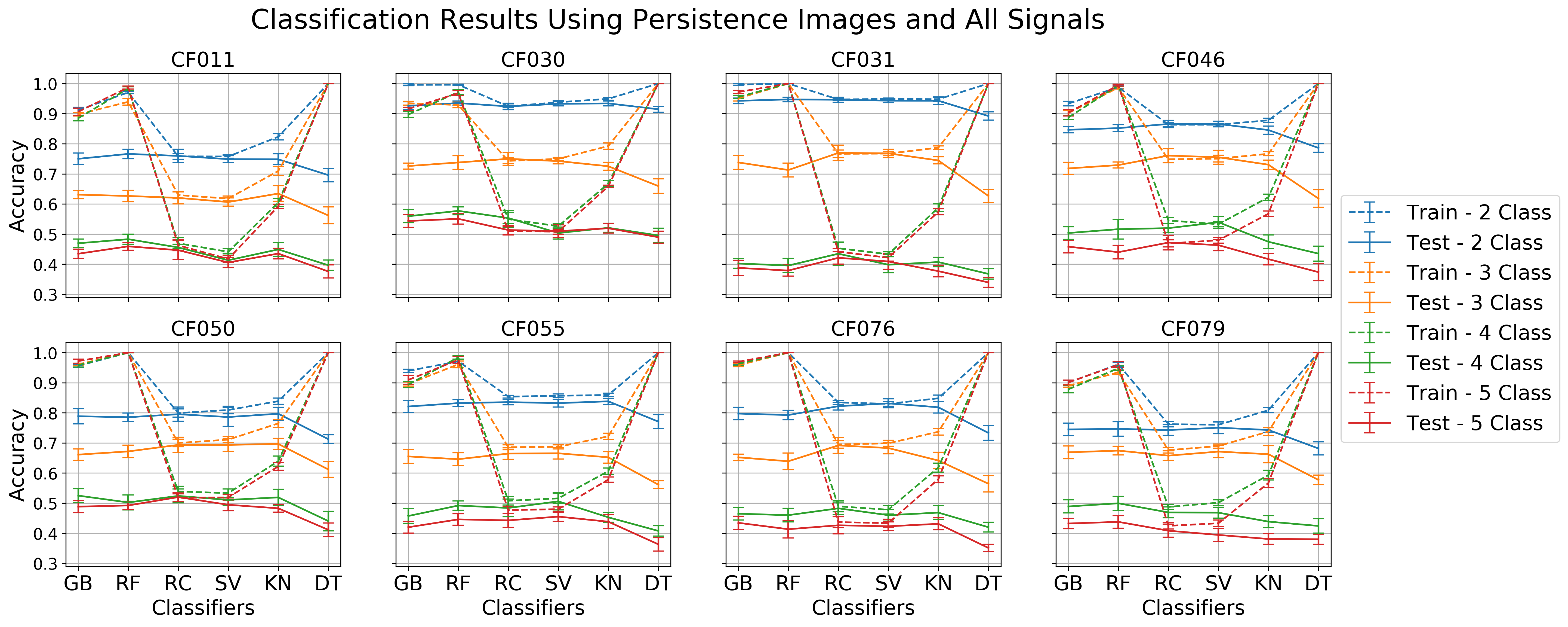}
    \caption{Average classification accuracies using persistence images for 2, 3, 4 and 5 classes for each patient using all three signals. Error bars are the standard deviation. Classifiers used are gradient boosting (GB), random forest (RF), ridge classification (RC), support vector classifier (SV), K-neighbors classifier (KN) and decision tree (DT).}
    \label{fig:PI_ALL}
\end{figure}

 \begin{figure}
    \centering
    \includegraphics[width=0.9\textwidth]{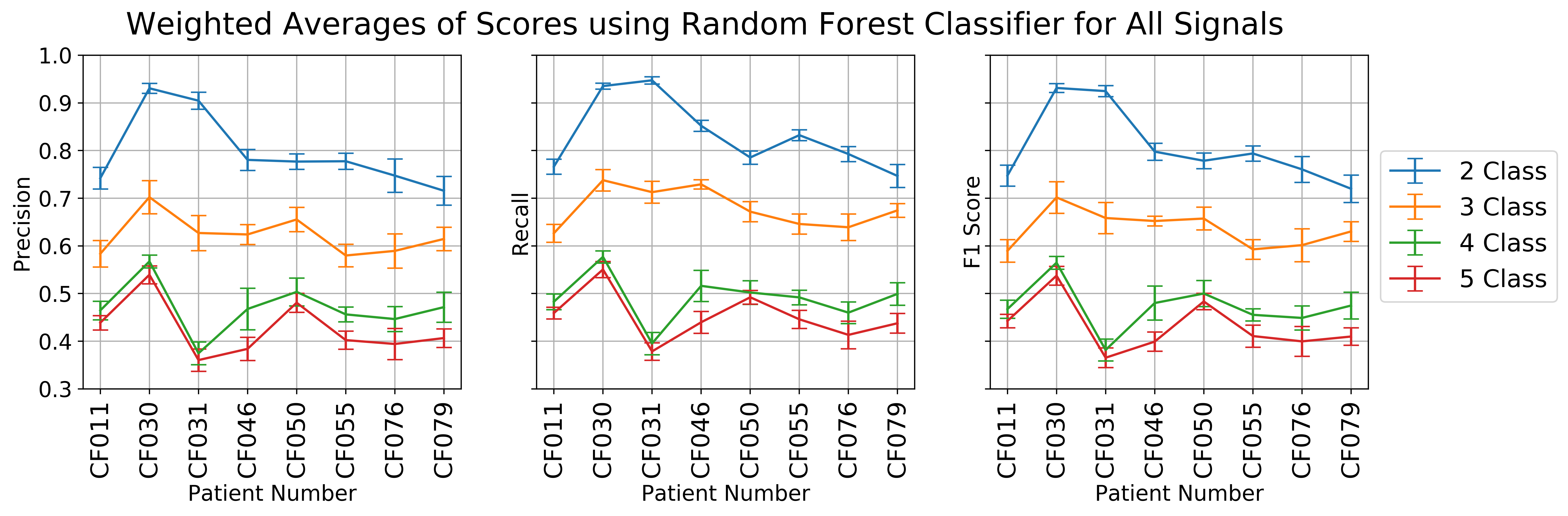}
    \caption{Weighted average precision, recall and F1 scores for Random Forest classification for each patient using 2, 3, 4 and 5 classes. Error bars are the standard deviation.}
    \label{fig:PI_F1}
\end{figure}

To apply the methods described in Sec.~\ref{ssec:PH_Background}, for each patient, we break the time series into non-overlapping 30 second intervals corresponding to the labeled epochs. 
Then for a given time series, we embed each 30 second interval separately using the Takens' embedding method and compute 1-dimensional persistence on the resulting point cloud.
For each persistence diagram we create a feature vector using persistence images.
Note we're only using 1-dimensional persistence diagrams because a time series with any periodic behavior will create a circular shape in the embedded point cloud.
An example of this transformation from persistence diagram to feature vector can be seen in Fig.~\ref{fig:pd_pi}. 
We will test each of the three PSG channels for classification separately, in addition to testing all of them combined.
We can test all of the channels together by computing their feature vectors with persistence images, and then concatenating the corresponding vectors for each epoch.
Note that for our analysis, we keep the data separated by patient in order to compare based on \emph{ahi} index as well.

For classification, we will test 2, 3, 4 and 5 class classification. 
The possible labels for each task are listed in Table~\ref{tab:classes}.
To perform the classification, we use the python package scikit-learn \cite{scikit-learn} to use six supervised machine learning algorithms: gradient boosting \cite{friedman2001greedy}, random forest \cite{breiman2001random}, ridge regression \cite{hoerl1970ridge}, support vector machines \cite{suykens1999least}, k-nearest neighbors classifier \cite{goldberger2005neighbourhood}, and decision tree \cite{breiman2017classification}.
For all of these classifiers, we use the default parameters.
For each experiment, we reserve 33\% for testing data and use the remaining for training.
We also run each classification task 10 times and average the accuracies across all 10 runs. 
The average and standard deviation of the accuracies for all classification algorithms can be seen in Fig.~\ref{fig:PI} and Fig.~\ref{fig:PI_ALL}.
 
From these results, we note that in Fig.~\ref{fig:PI}, each of the signals provide similar results for each patient. 
Thus one signal does not seem to contain more information about the sleep state than the others.
This is further emphasized by the fact that the results in Fig.~\ref{fig:PI_ALL} for all signals combined do not improve upon results from looking at a single signal.
For all patients, all classifiers the training accuracies vary, with random forests performing above 90\% for all patients, however the testing accuracies are about the same for a given number of classes.
It is expected that the performance of the classifiers is worse for more classes.
It appears that the 4 and 5 class classifications achieve relatively similar accuracies, while there is a big improvement reducing it to 3 classes. 

Looking across patients, there are variations. 
In general, for patient CF030, the classification accuracies are generally higher than for other patients for all numbers of classes. 
Patient CF079 seems to have the worst classification accuracies.
However, these variations in classification accuracies do not seem to correlate with the \emph{ahi} index. 

Existing methods compared in \cite{Peker2016} almost all achieve over 90\% accuracies.
However none of the methods mentioned use topological features.
In direct comparison, the authors in \cite{Chung2019} use similar persistent homology based approaches, however they use a different featurization method and they test their method across many databases.
They perform three different classification tasks: (1) sleep vs. wake, (2) REM vs. NREM, and (3) wake vs. REM vs. NREM.
For the first task, they report mean accuracies ranging from approximately 62-75\% accuracy with across datasets.
Their lowest reported accuracy, $62.6 \pm 17.0\%$, is from a dataset that contains patients with OSA.
Our accuracies for the two class classification problem are over 70\% for all patients. 
We acknowledge that we are only using 8 patients, while the datasets used in \cite{Chung2019} are much larger, so it is not a direct comparison.
However, the consistency of our results across these 8 patients with varying OSA severity seems promising that it would extend to larger datasets.

As shown in Table.~\ref{tab:sleepcycle}, we have very imbalanced data, and thus reporting accuracy alone may not fully report the results. 
Thus, in Fig.~\ref{fig:PI_F1} we report the weighted average for precision, recall and F1 score.
Each of these scores is defined based on the number of true positives (TP), true negatives (TN), false positives (FP) and false negatives (FN):
\[
\text{Precision} = \frac{TP}{TP + FP}, ~~~~ \text{Recall} = \frac{TP}{TP + FN}, ~~~~ \text{F1} = \frac{2\cdot TP}{2\cdot TP + FP + FN} 
\]
Each score is calculated for each class, then a weighted average is taken based on the number of samples with that class label. 
Note that if any class has no predicted samples, the score for that class is set to 0.
These calculations are done with the scikit-learn implementations.
We then take that weighted average for the 10 runs and take the average and standard deviation of those 10 values.

\section{Visualizing Sleep Patterns of Eight OSA Patients}\label{sec:markov}
In this section, our objective is to understand and visualize sleep patterns of the eight patients in our data set. As we see in Figure \ref{fig:combinedhypnogram}, a hypnogram records the sleep state at each epoch over the entire sleep duration of a patient. 
In our data, since the patients sleep for different time durations, we have different number of total epochs for each patient. However, we want to compare the hypnograms epoch by epoch. Therefore, we normalize them by truncating at the minimum number of epochs of a patient. We can make some observations just by looking at the hypnograms, one being that the sleep pattern of patient CF046 looks very different from the others. Similarly, we observe that the initial sleep stages for patient CF050 are different from the rest of the patients, see Figure \ref{fig:all_hypnograms} in the appendix, where the hypnograms of the remaining patients are depicted. 
These observations therefore lead to the conjecture that there is a relation between the sleep stage data and the severity of OSA in a patient. Therefore, we plan to use various statistical tools to study the sleep stage data and based on our results, determine the pairs of patients with same severity of OSA. This observation was made by one of authors, KS, who was blinded to the information of patients' OSA severity. 

We consider three  measures to inspect sleep patterns, namely transition probabilities, Cohen's $\kappa$, and Kullback-Leibler divergence. To do so,
we model sleep as a discrete Markov chain with 5 states. Let $(X_t)_{\mathbb{N}}$ denote the stationary categorical process with state space  $S=\{1, 2, 3, 4, 5\},$ corresponding to the sleep states \textbf{Wake}(W), \textbf{REM}(R), \textbf{NREM1}(N1), \textbf{NREM2}(N2), \textbf{NREM3}(N3), and  $\mathbf{\pi}=(\pi_{i_i}, \ldots, \pi_{i_5}),$ denote the marginal distribution for each patient $i.$ Table \ref{tab:margdis} shows the marginal distribution for the eight patients.

\begin{table}[h]
    \centering
    \begin{tabular}{|c|c|c|}
    \hline
    ~Patient ~&~ Marginal Distribution of (W, R, N1, N2, N3) ~&~ Order of frequencies \\
    \hline
  ~CF011  ~&~ $(0.247,0.146,0.023,{\bf 0.397},0.184)$ ~&~ $N2 > W > N3 > R > N1$~ \\
  \hline
  ~CF030 ~&~ $(0.078,0.192,0.013,{\bf 0.415},0.300)$ ~&~ $N2 > N3 > R > W > N1$~ \\
  \hline
  ~CF031 ~&~ $(0.052,0.183,0.030, 0.359, {\bf 0.375})$ ~&~ $N3 > N2 > R > W > N1$ \\
  \hline
  ~CF046 ~&~ $(0.133,0.110,0.086,0.319,{\bf 0.349})$ ~&~ $N3 > N2 > W > R > N1$ \\
  \hline
  ~CF050 ~&~ $(0.270,0.102,0.035, {\bf 0.321},0.269$ ~&~ $N2 > W > N3 > R > N1$ \\
  \hline
  ~CF055 ~&~ $(0.148,0.188,0.043,{\bf 0.411},0.208)$ ~&~ $N2 > N3 > R > W > N1$ \\
  \hline
  ~CF076 ~&~ $(0.170,0.135,0.050,{\bf 0.420},0.223)$ ~&~ $N2>N3>W>R>N1$ \\
  \hline
  ~CF079 ~&~ $(0.243,0.083,0.072,{\bf 0.407},0.192)$ ~&~ $N2>W>N3>R>N1$ \\
  \hline
    \end{tabular}
    \caption{Marginal Distributions $\pi$ of all $8$ patients}
    \label{tab:margdis}
\end{table}

For time-homogeneous Markov chain $X_t, t\in \mathbb{N},$ the transition probability is defined as 
$P(X_t=j|X_{t-1}=i):=p_{j|i}$ for for any $i, j \in S$ and all $t \in \mathbb{N}.$ We note that the sum of probabilities in each row equals to one, i.e.~$\sum_{j\in S} p_{j|i}=1$. The transition probability matrix and a visual representation of the 5-state Markov chain of a patient is depicted in Figures \ref{tab:tranpbty} and \ref{fig:tranPbty}.  

\begin{minipage}{0.4\textwidth}
\centering
\kbordermatrix{ & $W$ & $R$ & $N1$ & $N2$ & $N3$ \cr
    $W$ & 0.85 & 0.004 & 0.104 & 0.04 & 0 \cr
    $R$ &   0.082 & 0.882 & 0.035 & 0 & 0 \cr
    $N1$ &   0.094 & 0.094 & 0.54 & 0.27 & 0 \cr
    $N2$ &   0.048 & 0.004 & 0.012 & 0.925 & 0.009 \cr
    $N3$ &  0.015 & 0 & 0 & 0.005 & 0.979}
\captionof{figure}{Transition probability of CF079.}
\label{tab:tranpbty}
\end{minipage} \hfill
\begin{minipage}{0.49\textwidth}
\centering
     \scalebox{0.4}{
\begin{tikzpicture}
\node[circle,draw,minimum size = 20pt] (a) at (0,0) {Wake};
\node[circle,draw] (b) at (-4,-3) {REM};
\node[circle,draw] (c) at (-2,-6) {NREM1};
\node[circle,draw] (d) at (2,-6) {NREM2};
\node[circle,draw] (e) at (4,-3) {NREM3}; 
\path[->] (a) edge [loop above] node[blue] {0.85} (a);
\path[->] (a) edge node[left,blue]{0.004} (b);
\path[->] (a) edge node[right,blue]{0.104} (c);
\path[->] (a) edge node[right,blue]{0.04} (d);
\path[bend left,->] (b) edge node[left,blue]{0.082} (a);
\path (b) edge [loop left] node[blue] {0.882} (b);
\path[->] (b) edge node[right,blue]{0.035} (c);
\path[bend left,->] (c) edge node[left,blue]{0.094} (a);
\path[bend left,->] (c) edge node[left,blue]{0.094} (b);
\path[->] (c) edge [loop left] node[blue] {0.54} (c);
\path[->] (c) edge node[below,blue]{0.27} (d);
\path[bend right,->] (d) edge node[right,blue]{0.048} (a);
\path[bend right,->] (d) edge node[left,blue]{0.004} (b);
\path[bend left,->] (d) edge node[below,blue]{0.012} (c);
\path (d) edge [loop below] node[right,blue]{0.925} (d);
\path[->] (d) edge node[left,blue]{0.009} (e);
\path[bend right,->] (e) edge node[right,blue]{0.015} (a);
\path[bend left,->] (e) edge node[right,blue]{0.005} (d);
\path[->] (e) edge [loop right] node[right,blue]{0.979} (e);
\end{tikzpicture}

}
 
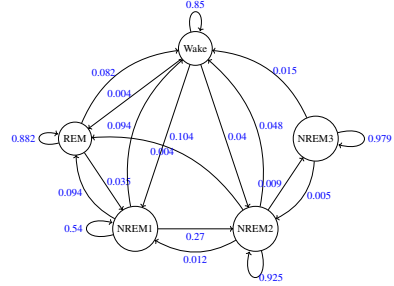
\captionof{figure}{Visual representation of Markov chain~(CF079).} \label{fig:tranPbty}
 \end{minipage}
\vskip0.3cm\noindent

To analyse the serial dependence structure of sleep state, we calculate Cohen's $\kappa$ at each lag $k$ as recommended by \cite{weiss-book}. Cohen's $\kappa$ is analogous to auto-correlation function in real-valued process (continuous time series). It is defined as
\[\kappa(k)=\cfrac{\sum_{j \in S}(P_{jj}(k)-\pi_j^2)}{1- \sum_{j \in S} \pi_j^2}, \]
where $P_{ij}(k)= P(X_t=i,\, X_{t-k}=j)$. 
The range of Cohen's $\kappa$ is $\left[- \sum_j\pi_j^2/
(1-\sum_j \pi_j^2), 1 \right]$. A positive (negative) value of $\kappa(k)$ means positive (negative) serial dependence, and a value of $0$ means serial independence at lag $k$. The Cohen's $\kappa(k)$ plots for $k=1, \ldots, 130$, are depicted for two patients in Figure \ref{fig:cohenFor2}, and for all patients in Figures \ref{fig:cohenAll} and \ref{fig:TPall} of the appendix. 

\begin{figure}
   \begin{tabular}{cccc}
       \scalebox{0.1}{\includegraphics{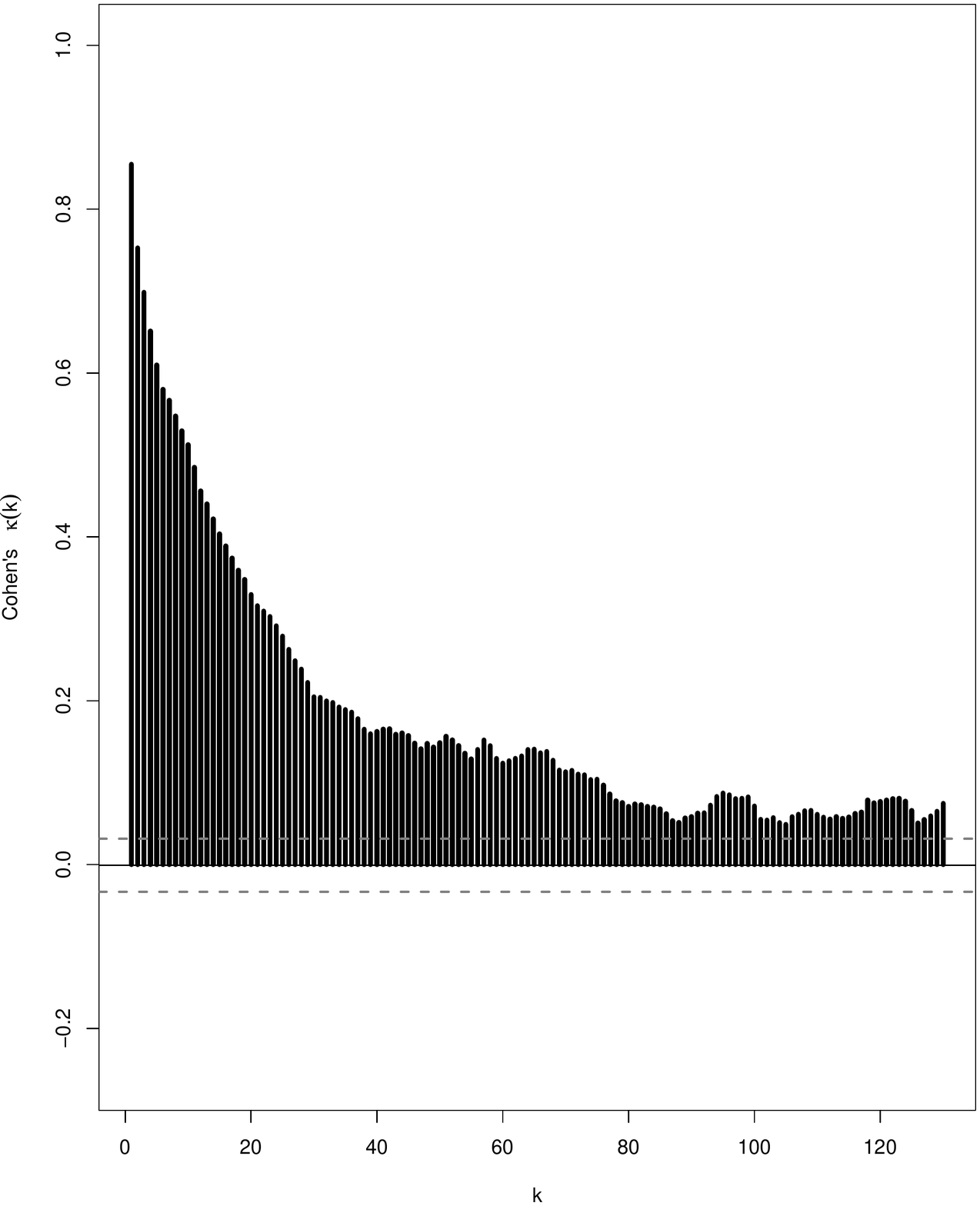}}~~ &\scalebox{0.3}{
\begin{tikzpicture}
\node[circle,draw,minimum size = 20pt] (a) at (0,0) {Wake};
\node[circle,draw] (b) at (-4,-3) {REM};
\node[circle,draw] (c) at (-2,-6) {NREM1};
\node[circle,draw] (d) at (2,-6) {NREM2};
\node[circle,draw] (e) at (4,-3) {NREM3}; 
\path[->] (a) edge [loop above] node[blue] {0.71} (a);
\path[->] (a) edge node[right,blue]{0.18} (c);
\path[->] (a) edge node[right,blue]{0.09} (d);
\path (b) edge [loop left] node[blue] {0.96} (b);
\path[->] (b) edge node[left,blue]{0.02} (c);
\path[->] (b) edge node[right,blue]{0.01} (d);
\path[bend left,->] (c) edge node[left,blue]{0.12} (a);
\path[bend left,->] (c) edge node[left,blue]{0.02} (b);
\path[->] (c) edge [loop left] node[blue] {0.66} (c);
\path[->] (c) edge node[below,blue]{0.2} (d);
\path[bend right,->] (d) edge node[right,blue]{0.08} (a);
\path (d) edge [loop below] node[right,blue]{0.9} (d);
\path[->] (d) edge node[left,blue]{0.01} (e);
\path[->] (e) edge [loop right] node[right,blue]{0.99} (e);
\end{tikzpicture}

}~~
      & \scalebox{0.1}{\includegraphics{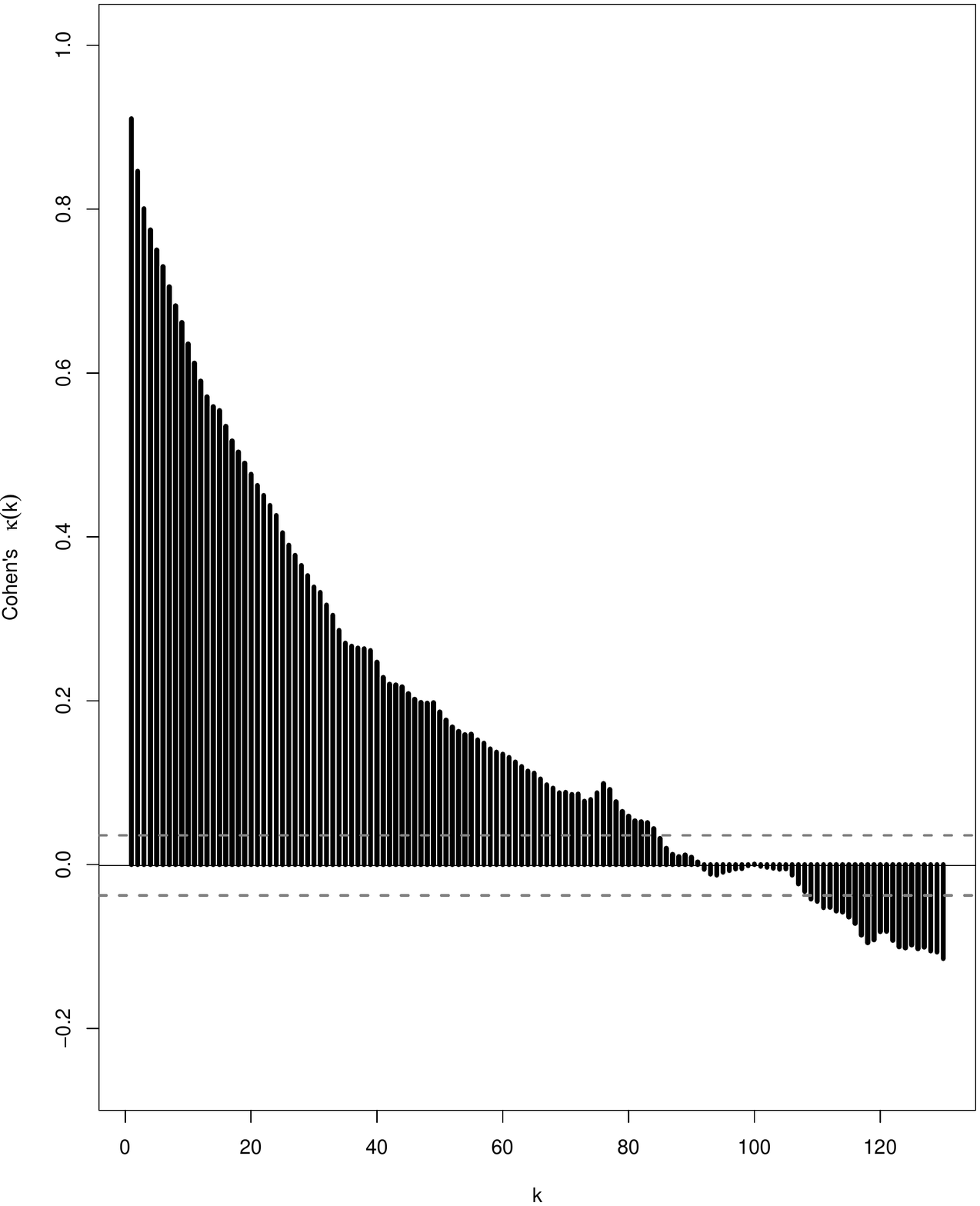}}~~
      &\scalebox{0.3}{
\begin{tikzpicture}
\node[circle,draw,minimum size = 20pt] (a) at (0,0) {Wake};
\node[circle,draw] (b) at (-4,-3) {REM};
\node[circle,draw] (c) at (-2,-6) {NREM1};
\node[circle,draw] (d) at (2,-6) {NREM2};
\node[circle,draw] (e) at (4,-3) {NREM3}; 
\path[->] (a) edge [loop above] node[blue] {0.93} (a);
\path[->] (a) edge node[right,blue]{0.06} (c);
\path (b) edge [loop left] node[blue] {0.98} (b);
\path[->] (b) edge node[right,blue]{0.02} (d);
\path[bend left,->] (c) edge node[left,blue]{0.11} (a);
\path[->] (c) edge [loop left] node[blue] {0.41} (c);
\path[->] (c) edge node[below,blue]{0.47} (d);
\path[bend right,->] (d) edge node[right,blue]{0.04} (a);
\path[bend left,->] (d) edge node[below,blue]{0.01} (c);
\path (d) edge [loop below] node[right,blue]{0.93} (d);
\path[->] (d) edge node[left,blue]{0.01} (e);
\path[bend right,->] (e) edge node[right,blue]{0.004} (a);
\path[bend right,->] (e) edge node[above,blue]{0.004} (b);
\path[bend right,->] (e) edge node[right,blue]{0.008} (c);
\path[->] (e) edge [loop right] node[right,blue]{0.98} (e);
\end{tikzpicture}

}
   \end{tabular}
\caption{(Left) Cohen's $\kappa$ plots and representation of transition probabilities for patients CF046~(left) and CF050~(right).}  
\label{fig:cohenFor2}
\end{figure}

We observe that Cohen's $\kappa (k)$ of patient CF050 is positive and gradually decreases until $k=90$, and then decreases to $-0.2$ at lag $k=130,$ see Figure \ref{fig:cohenFor2}.
This can be interpreted as each state tends to be followed by itself (but correlation decreases gradually), and then tends to flow from state to state.
The Cohen's $\kappa(k)$  of CF046 is positive at every lag $k=1, \ldots, 130,$ which means that the patient's sleep state is followed by itself. However, hypnogram of CF046 indicates that the patient stays at NREM 3 for long period of time during first, second and the last sleep cycles, while there is frequent flow between sleep states during the middle sleep cycle. Although, hypnogram and $\kappa(k)$ look different for CF046 and CF050, in terms of OSA severity, they are somewhat similar in that CF046 is moderate, and CF050 is severe. 

We now compare transition probability matrices $\p^1, \ldots, \p^8$ of the eight patients by calculating the Kullback-Leibler~(KL) \cite{KLD1951} divergence between probability distributions in each corresponding column in $\p$ and $\p'.$ 
As an example, KL between first row in each $\p$ and $\p'$ is
${\rm D}_{kl} (\p_1 \, \|\, \p'_1)= \sum_i p_{i|1}\ln(p_{i|1}/p'_{i|1}).$
We now define Kullback-Leibler divergence between $\p$ and $\p'$ as
\[{\rm D}_{kl} (\p\,\|\,\p')= \sum_i\sum_{j} p_{j|i}\, \ln \frac{ p_{j|i}}{p'_{j|i}}.\]
The KL divergence is asymmetric, the reason being that ${\rm KL}(\p\, \|\, \p')\neq {\rm KL}(\p'\,\|\, \p).$ We present sum of the two, i.e.~${\rm KL}(\p\, \|\, \p') + {\rm KL}(\p'\,\|\, \p),$ in Figure \ref{tab:KL}.

\begin{figure}
    \centering
    \[
    \kbordermatrix{ &	$CF079$	&	$CF076$	&	$CF030$	&	$CF011$	&	$CF031$	&	$CF046$	&	$CF055$	&	$CF050$	\cr
$CF079$	&	0	&	1.41	&	1.59	&	1.84	&	0.91	&	1.02	&	1.08	&	{\bf 2.00}	\cr
$CF076$	&		&	0	&	1.50	&	0.82	&	1.14	&	0.77	&	0.61	&	0.82	\cr
$CF030$	&		&		&	0	&	{\bf 2.96}	&	1.28	&	1.27	&	0.56	&	1.35	\cr
$CF011$	&		&		&		&	0	&	0.89	&	{\bf 2.19}	&	1.23	&	1.56	\cr
$CF031$	&		&		&		&		&	0	&	1.34	&	0.85	&	1.94	\cr
$CF046$	&		&		&		&		&		&	0	&	0.57	&	1.52	\cr
$CF055$	&		&		&		&		&		&		&	0	&	0.78	\cr
$CF050$	&		&		&		&		&		&		&		&	0}
\]
    \caption{Kullback-Leibler divergence between 8 patients.
    OSA severity of CF079 \& CF076: no, CF030 \& CF011: mild, CF031 \& CF046: moderate, CF055 \& CF050: severe. }
    \label{tab:KL}
\end{figure}

Ideally, KL divergence would be higher between  no and severe patients. As we notice from the table in Figure \ref{tab:KL}, there is no clear pattern. The patient CF050  with highest \emph{ahi} has high divergence from the patient CF079 with the  lowest \emph{ahi} and it is similar to the patient, CF055 (high  \emph{ahi}) as well as  CF076 (low \emph{ahi}). 
In conclusion, transition probabilities, Cohen's $\kappa$, and KL-divergence were not able to distinguish  patients with low \emph{ahi} from those with high \emph{ahi}.

In general, sleep patterns change through the night. At the beginning of sleep, a majority of people are in states mostly N2 and N3, with sporadic periods of N1 and short R periods. As the night progresses, the period of N3 becomes shorter, while N1 and N2 remain similar with longer R period. 
Both CF046 and CF050 don't follow the typical sleep patterns during sleep. The patient CF046 starts with a good sleep with a long period of N3 but after sleep cycle 2, frequent wake ups unable the patient to hit deep sleep until the morning. The pattern of CF050 is somewhat opposite to that of CF046; first frequent wake ups without falling in deep sleep during the early cycles, and then hit N3 during the last sleep cycle.  Neither of these two patterns are considered as ideal sleep. 
As we might expect, the patterns of transition probabilities are different in sleep cycles of patients. See for example, the transition probabilities in each sleep cycle for two patients in Figures \ref{fig:TP1} and \ref{fig:TP2} in the appendix. 
We therefore observe that it would be more useful to incorporate sleep cycles while calculating Cohen's $\kappa$, transition probabilities, and KL divergences.

\section{Conclusion and Future Research}\label{sec:conclusion}

In this paper, we first use a persistent homology approach for time series analysis to classify sleep states from three different PSG channels.
Our classification accuracies range drastically based on the number of classes used.
Additionally, the F1 scores reveal the underlying issue of class imbalance, which should be taken into account when considering our accuracies.
In the future, we would like to further work on this class imbalance issue, possibly by subsampling the data to get relatively equal distribution of classes in order to determine if this method using persistent homology is worth pursuing further.
We'd also like to test to see if using other featurizations of persistence diagrams yield different results. 

We modeled sleep stages as 5-state discrete first-order  Markov chain and try to discern sleep patterns of eight OSA patients with different OSA severity using transition probabilities, Cohen's $\kappa$, and Kullback-Leibler divergence. These three measures were not able to discern patients with low \emph{ahi} from those with  high \emph{ahi}.
This was our naive attempt to  understand any relationship between pattern of sleep stages and severity of OSA. Our next research goal is to analyse our data incorporating the sleep cycles. Furthermore,  we plan to consider higher order Markov chain.

\begin{acknowledgement}
This work was started at the second Women in Data Science and Mathematics workshop (WiSDM 2) in summer 2019, at The Institute for Computational and Experimental Research in Mathematics~(ICERM), Brown University. We thank ICERM for the support.
The authors would like to thank their fellow group members, Brenda Praggastis, Melissa Stockman, Kaisa Taipale, Marilyn Vazquez, Sunny Wang, and Emily Winn. 
We'd especially like to thank Brenda Praggastis for her help preprocessing the time series data and setting up some of the code for the persistent homology analysis.
We also thank Mathieu Chalifour for discussion of polysomnography time series.
ST was partially funded by NSF grants DMS1800446 and CMMI-1800466. KS was partially funded by NSF grant DMS 1547357.
GH would like to thank the National Sciences and Engineering Research Council of Canada (NSERC DG 2016-05167),
Seed grant from Women and Children's Health Research Institute, Biomedical Research Award from American Association of Orthodontists Foundation, and the McIntyre Memorial fund from the School of Dentistry at the University of Alberta.

\end{acknowledgement}

\newpage
\section{Appendix}
\addcontentsline{toc}{section}{Appendix}
This section contains figures referred to in the main article.
\begin{figure}
  \begin{tabular}{cccc}
\scalebox{0.12}{\includegraphics{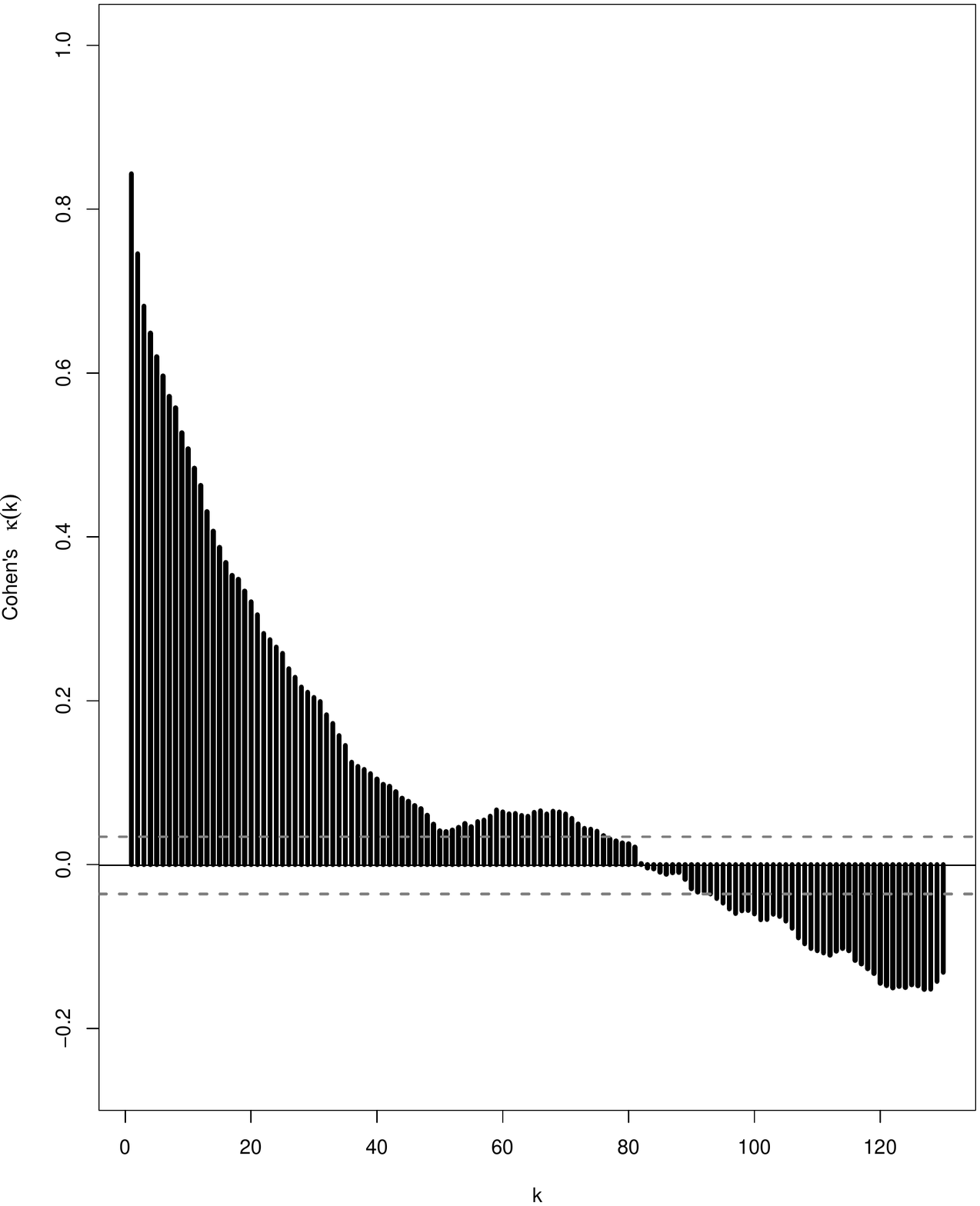}}~~
& \scalebox{0.12}{\includegraphics{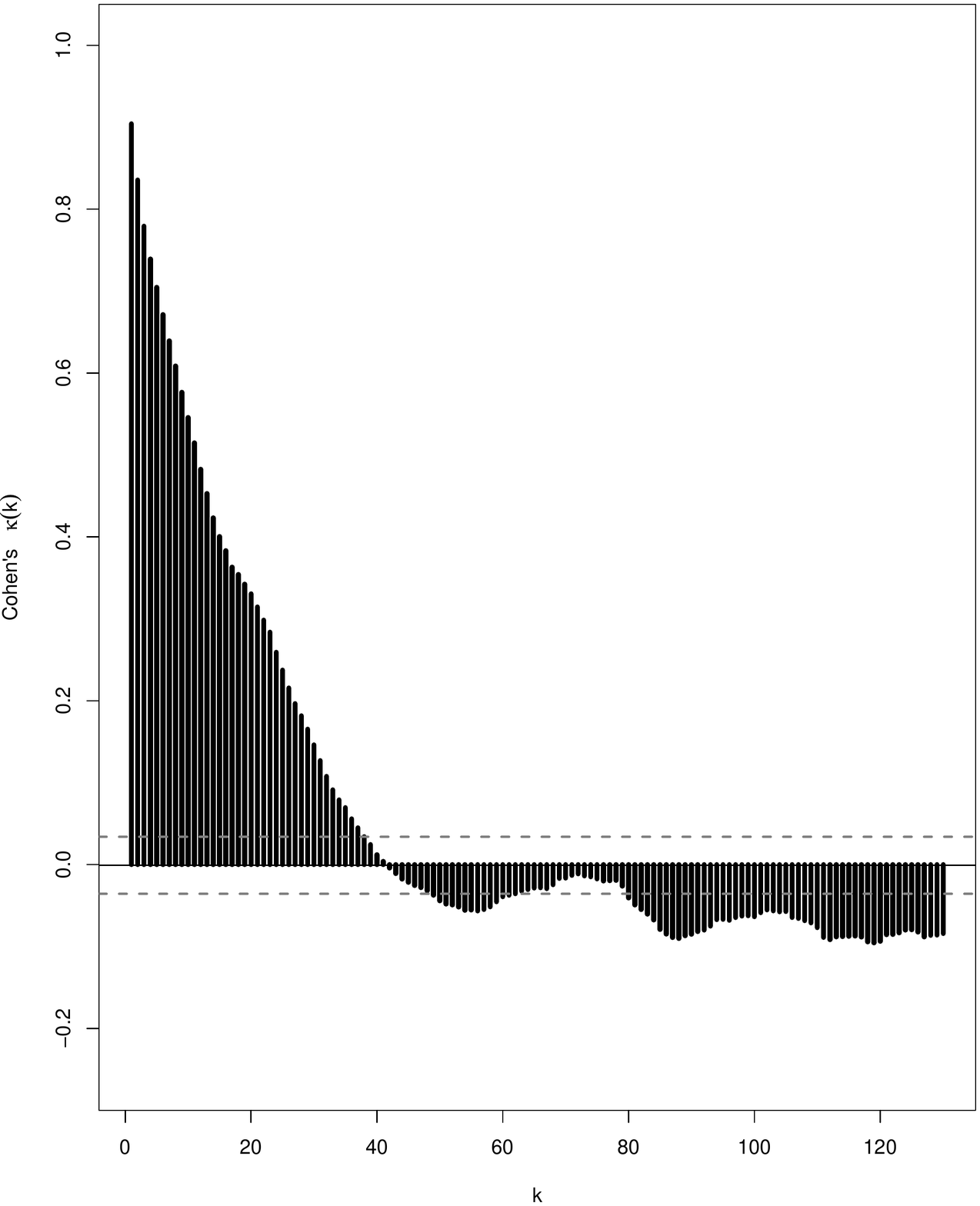}}~~
& \scalebox{0.12}{\includegraphics{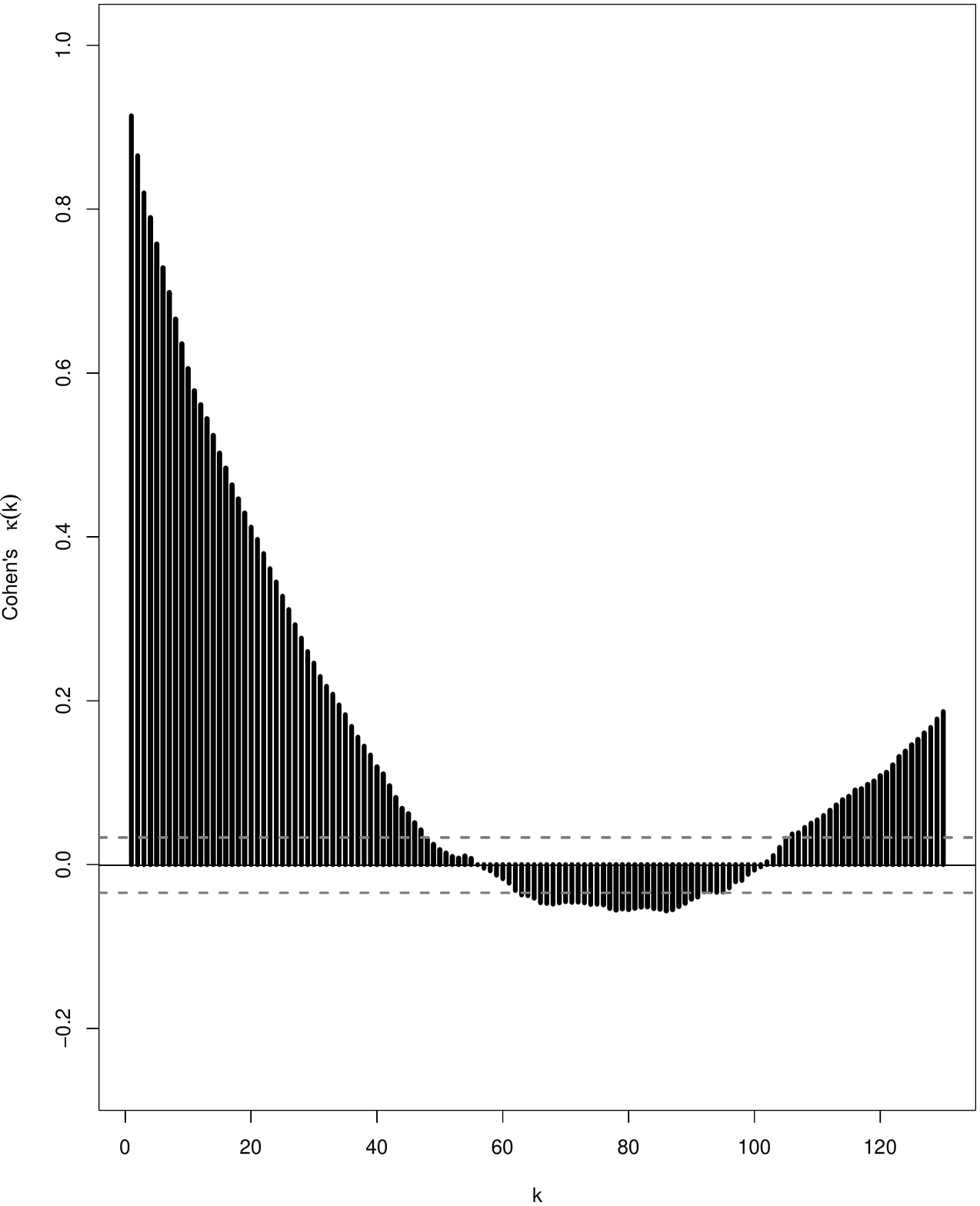}}~~
& \scalebox{0.12}{\includegraphics{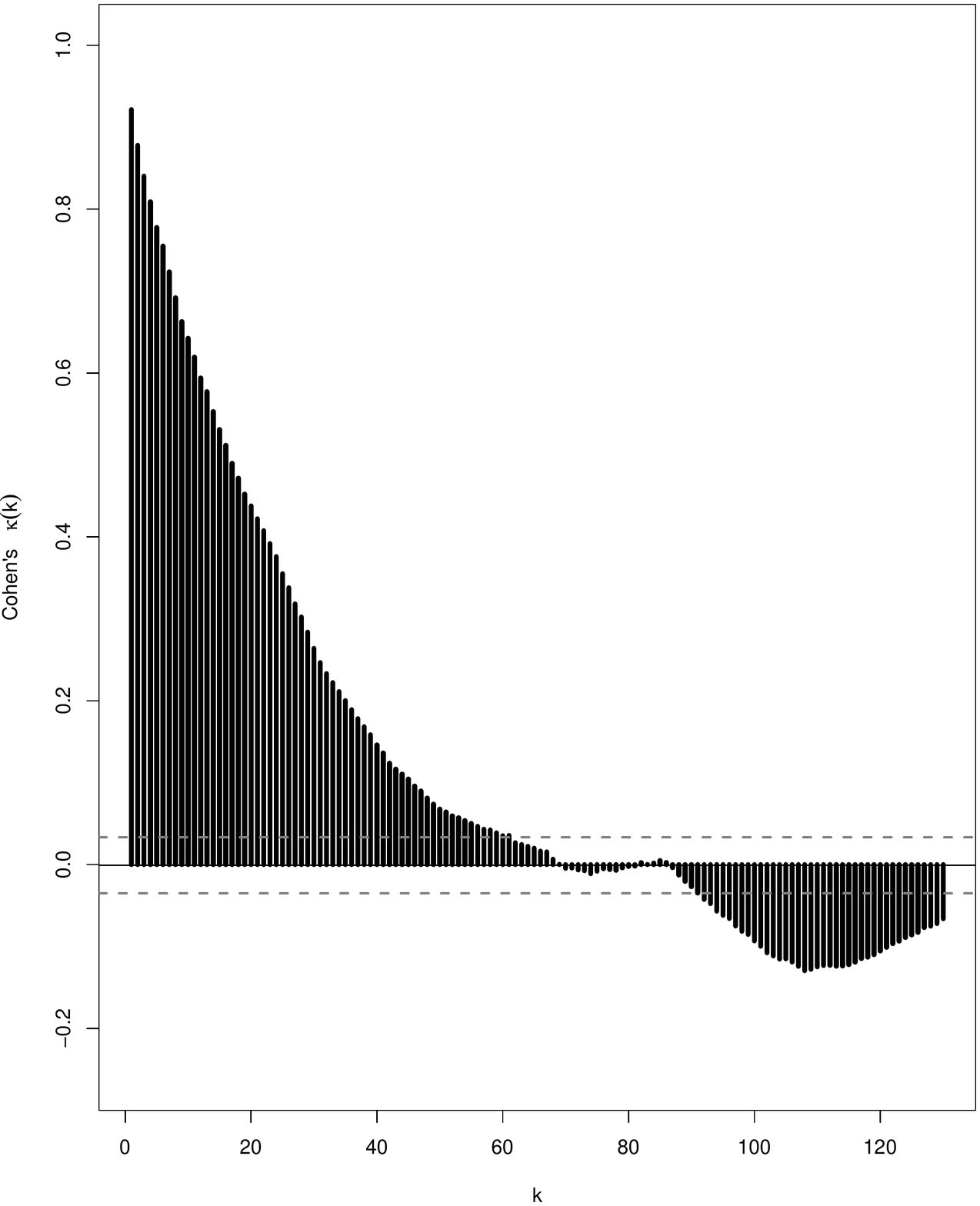}}\\
 \scalebox{0.12}{\includegraphics{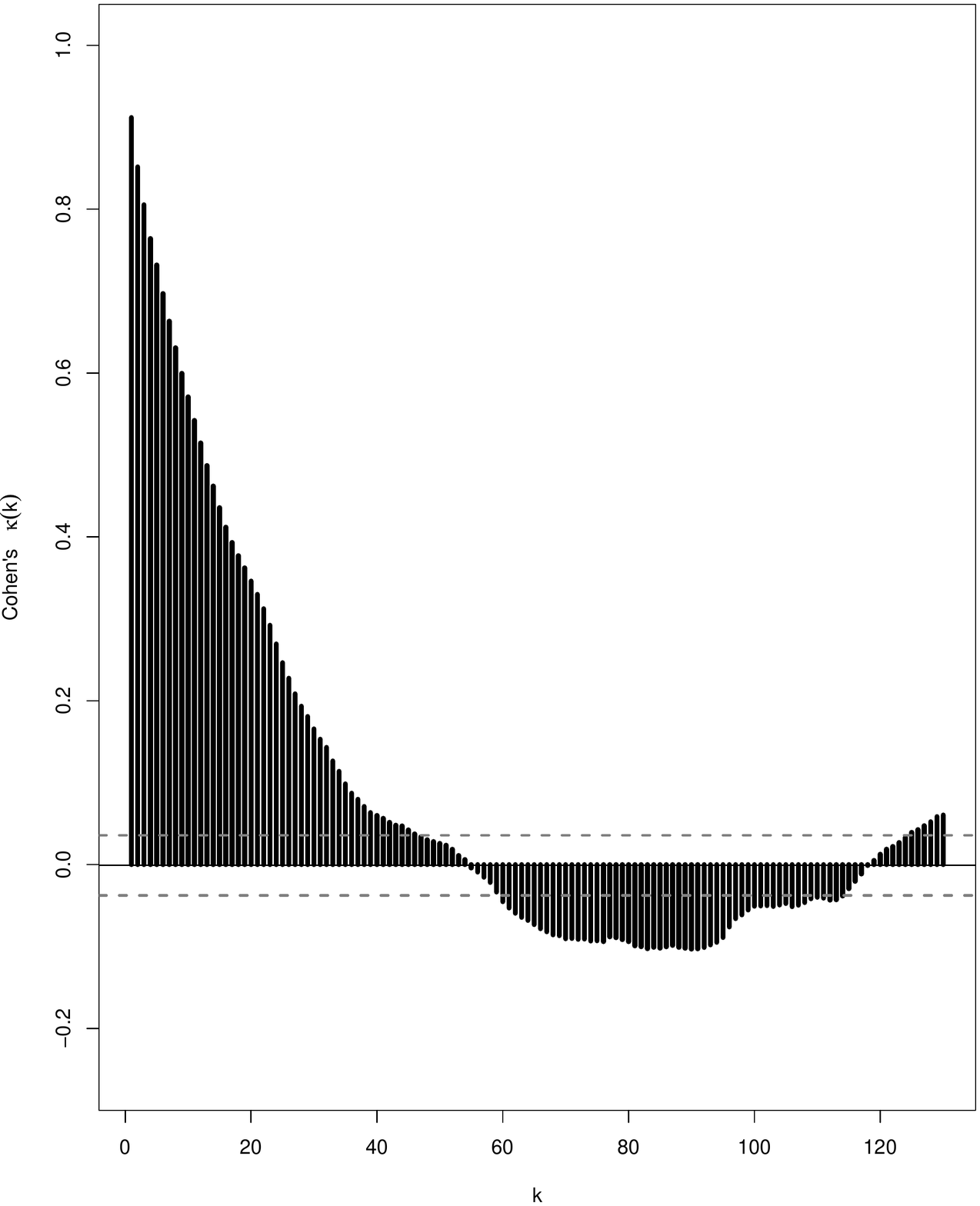}}~~  & \scalebox{0.12}{\includegraphics{Sleep_stage_figures/Data46_Cohen_Lag130.pdf}}~~
 & \scalebox{0.12}{\includegraphics{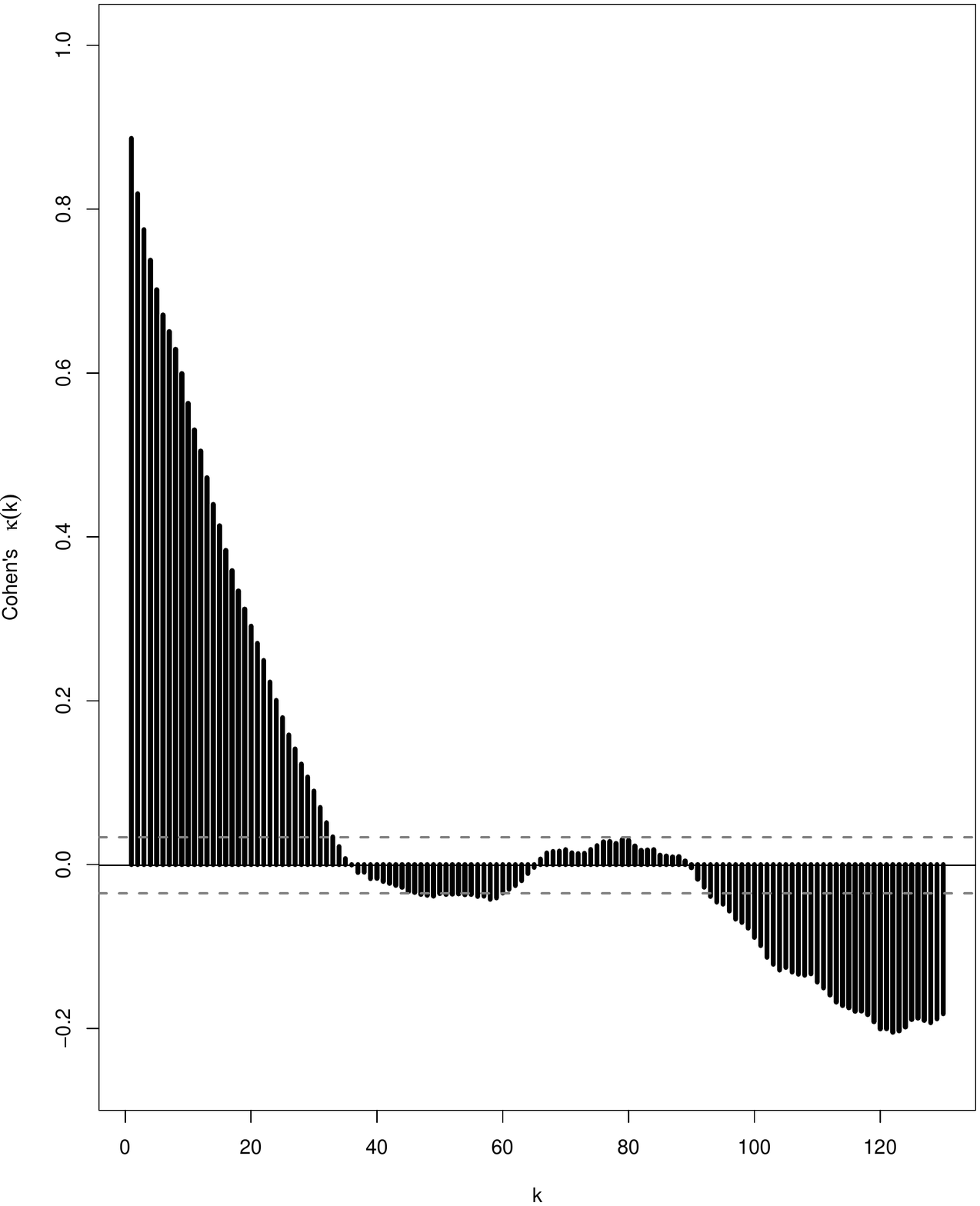}}~~
 & \scalebox{0.12}{\includegraphics{Sleep_stage_figures/Data50_Cohen_Lag130.pdf}}
  \end{tabular}
   \caption{Cohen's $\kappa$ for patients, CF079 \&CF076 (No OSA), CF030 \& CF011 (Mild OSA), CF031 \& CF046 (Moderate OSA), and CF055 \& CF050 (Severe OSA) (from left to right and top to bottom.)} 
   \label{fig:cohenAll}
\end{figure}

\begin{figure}
    \centering
    \scalebox{0.22}{\includegraphics{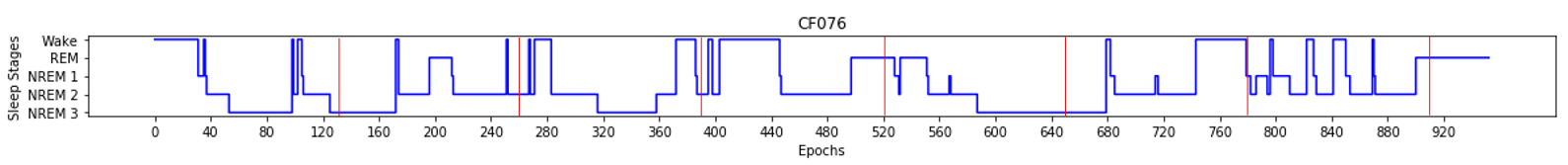}}
    \scalebox{0.22}{\includegraphics{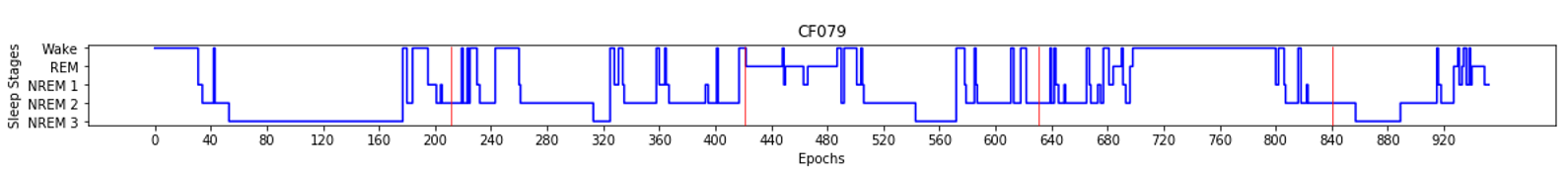}}
    \scalebox{0.22}{\includegraphics{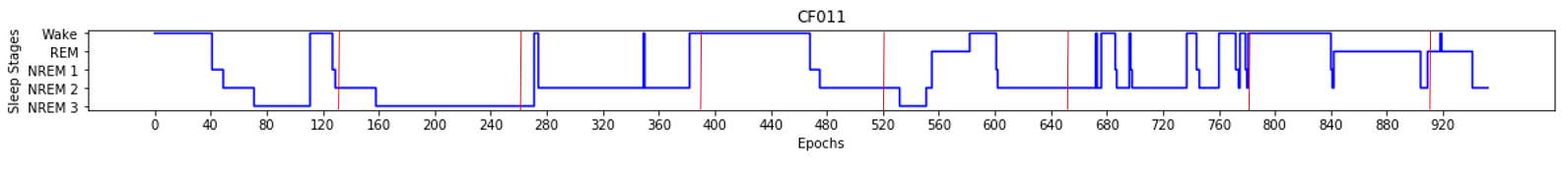}}
    \scalebox{0.22}{\includegraphics{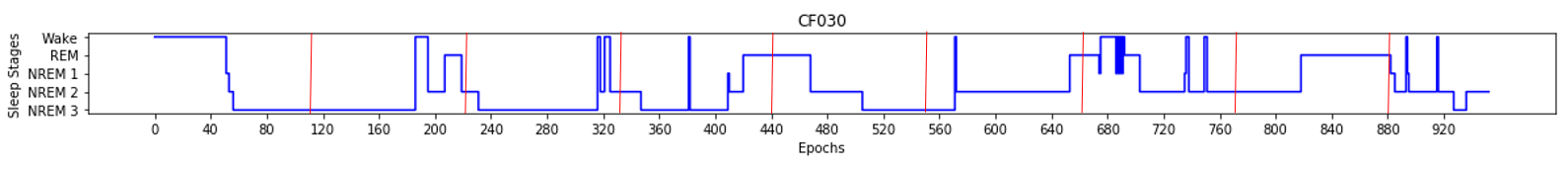}}
    \scalebox{0.22}{\includegraphics{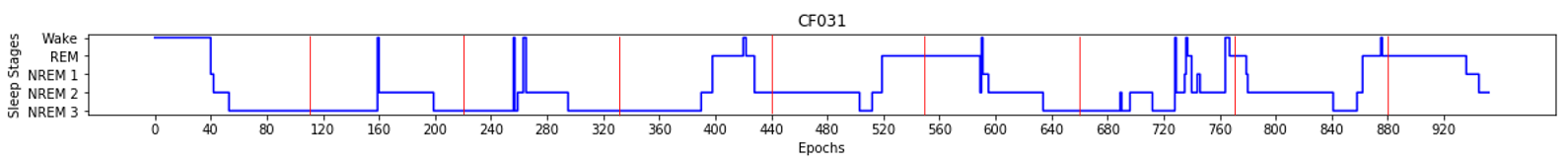}}
    \scalebox{0.22}{\includegraphics{Hypnograms/hypnogram_46_sleepcycle.png}}
    \scalebox{0.22}{\includegraphics{Hypnograms/hypnogram_50_sleepcycle.png}}
    \scalebox{0.22}{\includegraphics{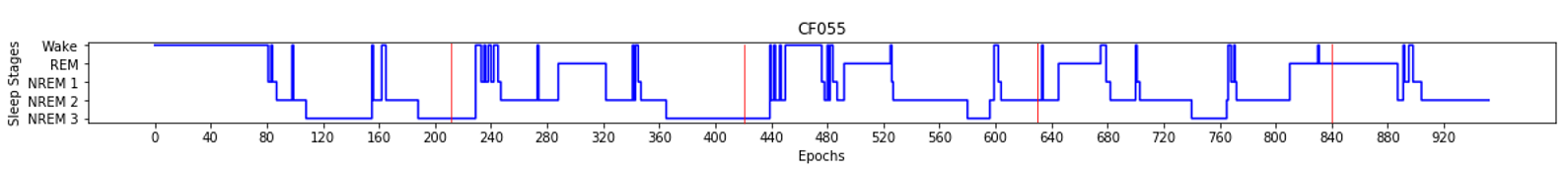}}
    \caption{Hypnograms of patients: CF079 \&CF076 (No OSA), CF030 \& CF011 (Mild OSA), CF031 \& CF046 (Moderate OSA), and CF055 \& CF050 (Severe OSA) with respective sleep cycles marked in red. 
    }
    \label{fig:all_hypnograms}
\end{figure}

\begin{figure}
    \begin{tabular}{cc}
       \scalebox{.5}{ 
\begin{tikzpicture}
\node[circle,draw,minimum size = 20pt] (a) at (0,0) {Wake};
\node[circle,draw] (b) at (-4,-3) {REM};
\node[circle,draw] (c) at (-2,-6) {NREM1};
\node[circle,draw] (d) at (2,-6) {NREM2};
\node[circle,draw] (e) at (4,-3) {NREM3}; 
\path[->] (a) edge [loop above] node[blue] {0.85} (a);
\path[->] (a) edge node[left,blue]{0.004} (b);
\path[->] (a) edge node[right,blue]{0.104} (c);
\path[->] (a) edge node[right,blue]{0.04} (d);
\path[bend left,->] (b) edge node[left,blue]{0.082} (a);
\path (b) edge [loop left] node[blue] {0.882} (b);
\path[->] (b) edge node[right,blue]{0.035} (c);
\path[bend left,->] (c) edge node[left,blue]{0.094} (a);
\path[bend left,->] (c) edge node[left,blue]{0.094} (b);
\path[->] (c) edge [loop left] node[blue] {0.54} (c);
\path[->] (c) edge node[below,blue]{0.27} (d);
\path[bend right,->] (d) edge node[right,blue]{0.048} (a);
\path[bend right,->] (d) edge node[left,blue]{0.004} (b);
\path[bend left,->] (d) edge node[below,blue]{0.012} (c);
\path (d) edge [loop below] node[right,blue]{0.925} (d);
\path[->] (d) edge node[left,blue]{0.009} (e);
\path[bend right,->] (e) edge node[right,blue]{0.015} (a);
\path[bend left,->] (e) edge node[right,blue]{0.005} (d);
\path[->] (e) edge [loop right] node[right,blue]{0.979} (e);
\end{tikzpicture}

} & \scalebox{0.5}{
\begin{tikzpicture}
\node[circle,draw,minimum size = 20pt] (a) at (0,0) {Wake};
\node[circle,draw] (b) at (-4,-3) {REM};
\node[circle,draw] (c) at (-2,-6) {NREM1};
\node[circle,draw] (d) at (2,-6) {NREM2};
\node[circle,draw] (e) at (4,-3) {NREM3}; 
\path[->] (a) edge [loop above] node[blue] {0.894} (a);
\path[->] (a) edge node[right,blue]{0.07} (c);
\path[->] (a) edge node[right,blue]{0.035} (d);
\path (b) edge [loop left] node[blue] {0.97} (b);
\path[->] (b) edge node[right,blue]{0.029} (c);
\path[bend left,->] (c) edge node[left,blue]{0.019} (a);
\path[->] (c) edge [loop left] node[blue] {0.627} (c);
\path[->] (c) edge node[below,blue]{0.352} (d);
\path[bend right,->] (d) edge node[right,blue]{0.032} (a);
\path[bend right,->] (d) edge node[left,blue]{0.009} (b);
\path[bend left,->] (d) edge node[below,blue]{0.007} (c);
\path (d) edge [loop below] node[right,blue]{0.941} (d);
\path[->] (d) edge node[left,blue]{0.009} (e);
\path[bend right,->] (e) edge node[right,blue]{0.013} (a);
\path[bend left,->] (e) edge node[right,blue]{0.004} (d);
\path[->] (e) edge [loop right] node[right,blue]{0.982} (e);
\end{tikzpicture}

} \\
         \scalebox{0.5}{
\begin{tikzpicture}
\node[circle,draw,minimum size = 20pt] (a) at (0,0) {Wake};
\node[circle,draw] (b) at (-4,-3) {REM};
\node[circle,draw] (c) at (-2,-6) {NREM1};
\node[circle,draw] (d) at (2,-6) {NREM2};
\node[circle,draw] (e) at (4,-3) {NREM3}; 
\path[->] (a) edge [loop above] node[blue] {0.78} (a);
\path[->] (a) edge node[left,blue]{0.03} (b);
\path[->] (a) edge node[right,blue]{0.07} (c);
\path[->] (a) edge node[right,blue]{0.11} (d);
\path[->] (a) edge node[left,blue]{0.01} (e);
\path[bend left,->] (b) edge node[left,blue]{0.01} (a);
\path (b) edge [loop left] node[blue] {0.96} (b);
\path[->] (b) edge node[left,blue]{0.008} (c);
\path[->] (b) edge node[right,blue]{0.016} (d);
\path[bend left,->] (c) edge node[left,blue]{0.3} (a);
\path[->] (c) edge [loop left] node[blue] {0.3} (c);
\path[->] (c) edge node[below,blue]{0.4} (d);
\path[bend right,->] (d) edge node[right,blue]{0.02} (a);
\path[bend right,->] (d) edge node[left,blue]{0.01} (b);
\path (d) edge [loop below] node[right,blue]{0.95} (d);
\path[->] (d) edge node[left,blue]{0.01} (e);
\path[bend right,->] (e) edge node[right,blue]{0.01} (a);
\path[bend left,->] (e) edge node[right,blue]{0.01} (d);
\path[->] (e) edge [loop right] node[right,blue]{0.97} (e);
\end{tikzpicture}

} &
        \scalebox{0.5}{
\begin{tikzpicture}
\node[circle,draw,minimum size = 20pt] (a) at (0,0) {Wake};
\node[circle,draw] (b) at (-4,-3) {REM};
\node[circle,draw] (c) at (-2,-6) {NREM1};
\node[circle,draw] (d) at (2,-6) {NREM2};
\node[circle,draw] (e) at (4,-3) {NREM3}; 
\path[->] (a) edge [loop above] node[blue] {0.92} (a);
\path[->] (a) edge node[left,blue]{0.02} (b);
\path[->] (a) edge node[right,blue]{0.03} (c);
\path[->] (a) edge node[right,blue]{0.02} (d);
\path[bend left,->] (b) edge node[left,blue]{0.05} (a);
\path (b) edge [loop left] node[blue] {0.94} (b);
\path[->] (b) edge node[right,blue]{0.01} (d);
\path (c) edge [loop left] node[blue] {0.61} (c);
\path[->] (c) edge node[below,blue]{0.38} (d);
\path[bend right,->] (d) edge node[right,blue]{0.02} (a);
\path[bend right,->] (d) edge node[left,blue]{0.01} (b);
\path (d) edge [loop below] node[right,blue]{0.96} (d);
\path[->] (d) edge node[right,blue]{0.01} (e);
\path[->] (e) edge node[right,blue]{0.015} (a);
\path[bend left,->] (e) edge node[right,blue]{0.005} (d);
\path[->] (e) edge [loop right] node[right,blue]{0.98} (e);
\end{tikzpicture}

} \\
\scalebox{0.5}{
\begin{tikzpicture}
\node[circle,draw,minimum size = 20pt] (a) at (0,0) {Wake};
\node[circle,draw] (b) at (-4,-3) {REM};
\node[circle,draw] (c) at (-2,-6) {NREM1};
\node[circle,draw] (d) at (2,-6) {NREM2};
\node[circle,draw] (e) at (4,-3) {NREM3}; 
\path[->] (a) edge [loop above] node[blue] {0.78} (a);
\path[->] (a) edge node[left,blue]{0.08} (b);
\path[->] (a) edge node[right,blue]{0.05} (c);
\path[->] (a) edge node[right,blue]{0.07} (d);
\path[->] (a) edge node[left,blue]{0.01} (e);
\path[bend left,->] (b) edge node[left,blue]{0.01} (a);
\path (b) edge [loop left] node[blue] {0.94} (b);
\path[->] (b) edge node[right,blue]{0.02} (c);
\path[->] (b) edge node[right,blue]{0.02} (d);
\path[bend left,->] (c) edge node[left,blue]{0.03} (a);
\path[bend left,->] (c) edge node[left,blue]{0.03} (b);
\path[->] (c) edge [loop left] node[blue] {0.67} (c);
\path[->] (c) edge node[below,blue]{0.26} (d);
\path[bend right,->] (d) edge node[right,blue]{0.01} (a);
\path[bend right,->] (d) edge node[left,blue]{0.01} (b);
\path[bend left,->] (d) edge node[below,blue]{0.01} (c);
\path (d) edge [loop below] node[right,blue]{0.94} (d);
\path[->] (d) edge node[left,blue]{0.02} (e);
\path[bend right,->] (e) edge node[right,blue]{0.01} (a);
\path[bend left,->] (e) edge node[right,blue]{0.01} (d);
\path[->] (e) edge [loop right] node[right,blue]{0.97} (e);
\end{tikzpicture}

} &
       \scalebox{0.5}{
\begin{tikzpicture}
\node[circle,draw,minimum size = 20pt] (a) at (0,0) {Wake};
\node[circle,draw] (b) at (-4,-3) {REM};
\node[circle,draw] (c) at (-2,-6) {NREM1};
\node[circle,draw] (d) at (2,-6) {NREM2};
\node[circle,draw] (e) at (4,-3) {NREM3}; 
\path[->] (a) edge [loop above] node[blue] {0.71} (a);
\path[->] (a) edge node[right,blue]{0.18} (c);
\path[->] (a) edge node[right,blue]{0.09} (d);
\path (b) edge [loop left] node[blue] {0.96} (b);
\path[->] (b) edge node[left,blue]{0.02} (c);
\path[->] (b) edge node[right,blue]{0.01} (d);
\path[bend left,->] (c) edge node[left,blue]{0.12} (a);
\path[bend left,->] (c) edge node[left,blue]{0.02} (b);
\path[->] (c) edge [loop left] node[blue] {0.66} (c);
\path[->] (c) edge node[below,blue]{0.2} (d);
\path[bend right,->] (d) edge node[right,blue]{0.08} (a);
\path (d) edge [loop below] node[right,blue]{0.9} (d);
\path[->] (d) edge node[left,blue]{0.01} (e);
\path[->] (e) edge [loop right] node[right,blue]{0.99} (e);
\end{tikzpicture}

} \\
         \scalebox{0.5}{
\begin{tikzpicture}
\node[circle,draw,minimum size = 20pt] (a) at (0,0) {Wake};
\node[circle,draw] (b) at (-4,-3) {REM};
\node[circle,draw] (c) at (-2,-6) {NREM1};
\node[circle,draw] (d) at (2,-6) {NREM2};
\node[circle,draw] (e) at (4,-3) {NREM3}; 
\path[->] (a) edge [loop above] node[blue] {0.813} (a);
\path[->] (a) edge node[left,blue]{0.006} (b);
\path[->] (a) edge node[right,blue]{0.116} (c);
\path[->] (a) edge node[right,blue]{0.064} (d);
\path[bend left,->] (b) edge node[left,blue]{0.015} (a);
\path (b) edge [loop left] node[blue] {0.974} (b);
\path[->] (b) edge node[right,blue]{0.01} (d);
\path[bend left,->] (c) edge node[left,blue]{0.133} (a);
\path[->] (c) edge [loop left] node[blue] {0.577} (c);
\path[->] (c) edge node[below,blue]{0.288} (d);
\path[bend right,->] (d) edge node[right,blue]{0.035} (a);
\path[bend right,->] (d) edge node[left,blue]{0.011} (b);
\path[bend left,->] (d) edge node[below,blue]{0.002} (c);
\path (d) edge [loop below] node[right,blue]{0.937} (d);
\path[->] (d) edge node[left,blue]{0.014} (e);
\path[bend right,->] (e) edge node[right,blue]{0.018} (a);
\path[bend left,->] (e) edge node[right,blue]{0.009} (d);
\path[->] (e) edge [loop right] node[right,blue]{0.972} (e);
\end{tikzpicture}

} & \scalebox{0.5}{
\begin{tikzpicture}
\node[circle,draw,minimum size = 20pt] (a) at (0,0) {Wake};
\node[circle,draw] (b) at (-4,-3) {REM};
\node[circle,draw] (c) at (-2,-6) {NREM1};
\node[circle,draw] (d) at (2,-6) {NREM2};
\node[circle,draw] (e) at (4,-3) {NREM3}; 
\path[->] (a) edge [loop above] node[blue] {0.93} (a);
\path[->] (a) edge node[right,blue]{0.06} (c);
\path (b) edge [loop left] node[blue] {0.98} (b);
\path[->] (b) edge node[right,blue]{0.02} (d);
\path[bend left,->] (c) edge node[left,blue]{0.11} (a);
\path[->] (c) edge [loop left] node[blue] {0.41} (c);
\path[->] (c) edge node[below,blue]{0.47} (d);
\path[bend right,->] (d) edge node[right,blue]{0.04} (a);
\path[bend left,->] (d) edge node[below,blue]{0.01} (c);
\path (d) edge [loop below] node[right,blue]{0.93} (d);
\path[->] (d) edge node[left,blue]{0.01} (e);
\path[bend right,->] (e) edge node[right,blue]{0.004} (a);
\path[bend right,->] (e) edge node[above,blue]{0.004} (b);
\path[bend right,->] (e) edge node[right,blue]{0.008} (c);
\path[->] (e) edge [loop right] node[right,blue]{0.98} (e);
\end{tikzpicture}

}
    \end{tabular}
    \caption{Transition probabilities for patients CF079, CF076, CF030, CF031, CF046, CF055, and CF050 increasing order of \emph{ahi} scores (left to right, and top to bottom).}
    \label{fig:TPall}
\end{figure}
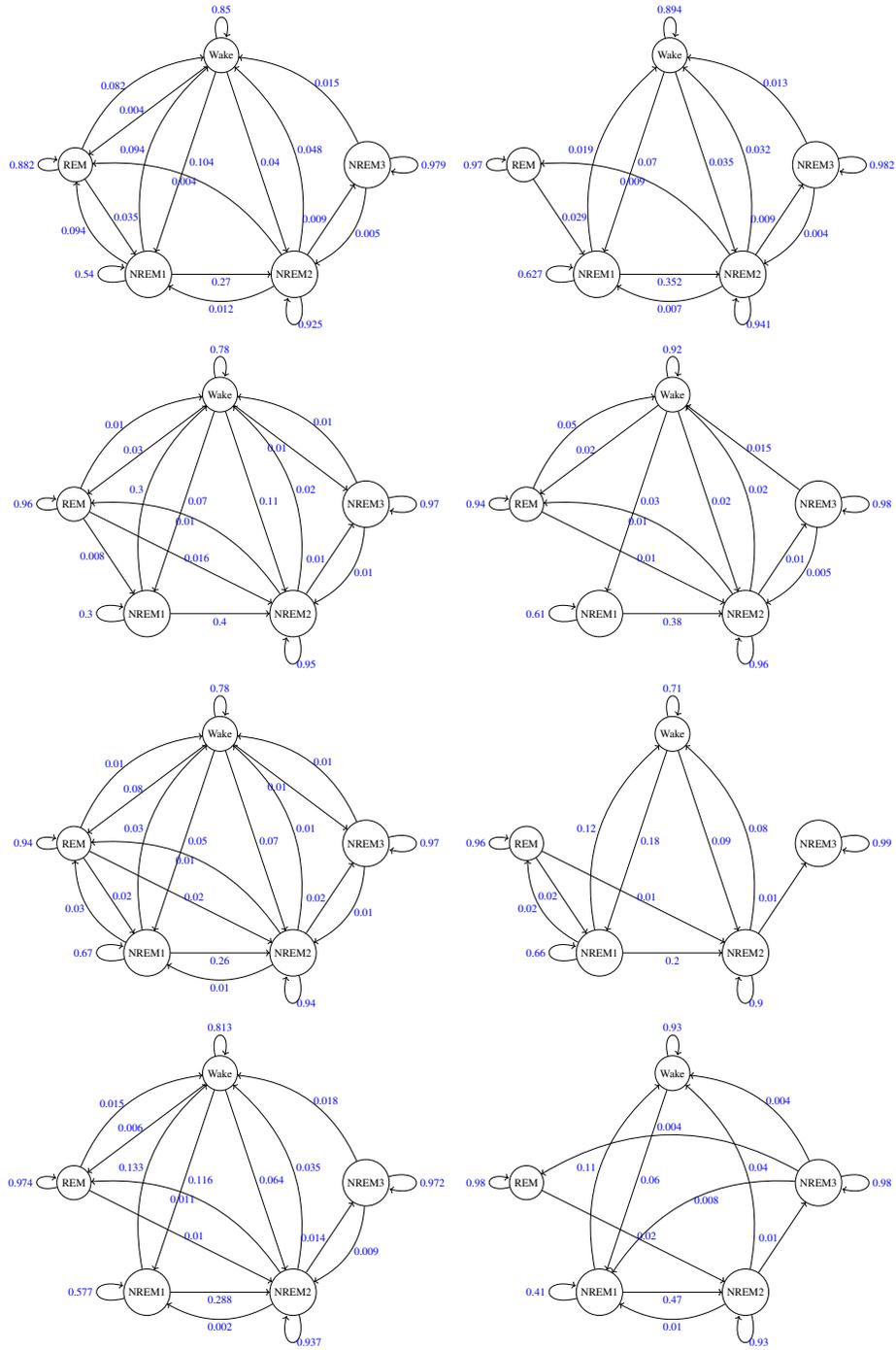

\begin{figure}
    \begin{tabular}{cc}
\scalebox{.5}{
\begin{tikzpicture}
\node[circle,draw,minimum size = 20pt] (a) at (0,0) {Wake};
\node[circle,draw] (c) at (-2,-6) {NREM1};
\node[circle,draw] (d) at (2,-6) {NREM2};
\node[circle,draw] (e) at (4,-3) {NREM3}; 
\path[->] (a) edge [loop above] node[blue] {0.96} (a);
\path[->] (a) edge node[right,blue]{0.04} (c);
\path[->] (c) edge [loop left] node[blue] {0.5} (c);
\path[->] (c) edge node[below,blue]{0.5} (d);
\path (d) edge [loop below] node[right,blue]{0.9} (d);
\path[->] (d) edge node[left,blue]{0.1} (e);
\path[->] (e) edge [loop right] node[right,blue]{1.00} (e);
\end{tikzpicture}
}
 &     
\scalebox{.5}{
\begin{tikzpicture}
\node[circle,draw,minimum size = 20pt] (a) at (0,0) {Wake};
\node[circle,draw] (d) at (2,-6) {NREM2};
\node[circle,draw] (e) at (4,-3) {NREM3}; 
\path[->] (a) edge node[right,blue]{1.00} (d);
\path (d) edge [loop below] node[right,blue]{1.00} (d);
\path[bend right,->] (e) edge node[right,blue]{0.008} (a);
\path[->] (e) edge [loop right] node[right,blue]{0.991} (e);
\end{tikzpicture}
}
 \\
\scalebox{.5}{
\begin{tikzpicture}
\node[circle,draw,minimum size = 20pt] (a) at (0,0) {Wake};
\node[circle,draw] (b) at (-4,-3) {REM};
\node[circle,draw] (c) at (-2,-6) {NREM1};
\node[circle,draw] (d) at (2,-6) {NREM2};
\path[->] (a) edge [loop above] node[blue] {0.529} (a);
\path[->] (a) edge node[right,blue]{0.176} (c);
\path[->] (a) edge node[right,blue]{0.294} (d);
\path (b) edge [loop left] node[blue] {0.935} (b);
\path[->] (b) edge node[right,blue]{0.032} (c);
\path[->] (b) edge node[right,blue]{0.032} (d);
\path[bend left,->] (c) edge node[left,blue]{0.222} (a);
\path[bend left,->] (c) edge node[left,blue]{0.111} (b);
\path[->] (c) edge [loop left] node[blue] {0.555} (c);
\path[->] (c) edge node[below,blue]{0.111} (d);
\path[bend right,->] (d) edge node[right,blue]{0.097} (a);
\path[bend right,->] (d) edge node[left,blue]{0.013} (b);
\path (d) edge [loop below] node[right,blue]{0.888} (d);
\end{tikzpicture}
}
 &
\scalebox{.5}{
\begin{tikzpicture}
\node[circle,draw,minimum size = 20pt] (a) at (0,0) {Wake};
\node[circle,draw] (c) at (-2,-6) {NREM1};
\node[circle,draw] (d) at (2,-6) {NREM2};
\path[->] (a) edge [loop above] node[blue] {0.727} (a);
\path[->] (a) edge node[right,blue]{0.151} (c);
\path[->] (a) edge node[right,blue]{0.121} (d);
\path[bend left,->] (c) edge node[left,blue]{0.176} (a);
\path[->] (c) edge [loop left] node[blue] {0.705} (c);
\path[->] (c) edge node[below,blue]{0.117} (d);
\path[bend right,->] (d) edge node[right,blue]{0.063} (a);
\path (d) edge [loop below] node[right,blue]{0.936} (d);
\end{tikzpicture}
}
 \\
\scalebox{.5}{
\begin{tikzpicture}
\node[circle,draw,minimum size = 20pt] (a) at (0,0) {Wake};
\node[circle,draw] (b) at (-4,-3) {REM};
\node[circle,draw] (c) at (-2,-6) {NREM1};
\node[circle,draw] (d) at (2,-6) {NREM2};
\path[->] (a) edge [loop above] node[blue] {0.56} (a);
\path[->] (a) edge node[right,blue]{0.32} (c);
\path[->] (a) edge node[right,blue]{0.12} (d);
\path (b) edge [loop left] node[blue] {0.941} (b);
\path[->] (b) edge node[right,blue]{0.058} (c);
\path[bend left,->] (c) edge node[left,blue]{0.111} (a);
\path[bend left,->] (c) edge node[left,blue]{0.027} (b);
\path[->] (c) edge [loop left] node[blue] {0.722} (c);
\path[->] (c) edge node[below,blue]{0.138} (d);
\path[bend right,->] (d) edge node[right,blue]{0.137} (a);
\path[bend left,->] (d) edge node[below,blue]{0.019} (c);
\path (d) edge [loop below] node[right,blue]{0.843} (d);
\end{tikzpicture}
}
 & 
\scalebox{.5}{
\begin{tikzpicture}
\node[circle,draw,minimum size = 20pt] (a) at (0,0) {Wake};
\node[circle,draw] (c) at (-2,-6) {NREM1};
\node[circle,draw] (d) at (2,-6) {NREM2};
\path[->] (a) edge [loop above] node[blue] {0.777} (a);
\path[->] (a) edge node[right,blue]{0.2} (c);
\path[->] (a) edge node[right,blue]{0.022} (d);
\path[bend left,->] (c) edge node[left,blue]{0.09} (a);
\path[->] (c) edge [loop left] node[blue] {0.59} (c);
\path[->] (c) edge node[below,blue]{0.318} (d);
\path[bend right,->] (d) edge node[right,blue]{0.129} (a);
\path (d) edge [loop below] node[right,blue]{0.870} (d);
\end{tikzpicture}
}
 \\ 
\scalebox{.5}{
\begin{tikzpicture}
\node[circle,draw,minimum size = 20pt] (a) at (0,0) {Wake};
\node[circle,draw] (b) at (-4,-3) {REM};
\node[circle,draw] (c) at (-2,-6) {NREM1};
\node[circle,draw] (d) at (2,-6) {NREM2};
\node[circle,draw] (e) at (4,-3) {NREM3}; 
\path[->] (a) edge [loop above] node[blue] {0.5} (a);
\path[->] (a) edge node[right,blue]{0.333} (c);
\path[->] (a) edge node[right,blue]{0.166} (d);
\path (b) edge [loop left] node[blue] {1.00} (b);
\path[->] (c) edge [loop left] node[blue] {0.333} (c);
\path[->] (c) edge node[below,blue]{0.666} (d);
\path[bend right,->] (d) edge node[right,blue]{0.064} (a);
\path[bend right,->] (d) edge node[left,blue]{0.032} (b);
\path (d) edge [loop below] node[right,blue]{0.87} (d);
\path[->] (d) edge node[left,blue]{0.032} (e);
\path[bend right,->] (e) edge node[right,blue]{0.011} (a);
\path[->] (e) edge [loop right] node[right,blue]{0.988} (e);
\end{tikzpicture}
}
 & 
\scalebox{.5}{
\begin{tikzpicture}
\node[circle,draw] (b) at (-4,-3) {REM};
\node[circle,draw] (c) at (-2,-6) {NREM1};
\node[circle,draw] (d) at (2,-6) {NREM2};
\node[circle,draw] (e) at (4,-3) {NREM3}; 
\path (b) edge [loop left] node[blue] {0.979} (b);
\path[->] (b) edge node[right,blue]{0.02} (d);
\path[->] (c) edge node[below,blue]{1.00} (d);
\path (d) edge [loop below] node[right,blue]{0.976} (d);
\path[->] (d) edge node[left,blue]{0.023} (e);
\path[bend right,->] (e) edge node[right,blue]{0.027} (c);
\path[->] (e) edge [loop right] node[right,blue]{0.972} (e);
\end{tikzpicture}
}
    \end{tabular}
    \caption{Transition probabilities for the 8 sleep cycles of CF046 (displayed in order from left to right and top to bottom).} \label{fig:TP1}
\end{figure}
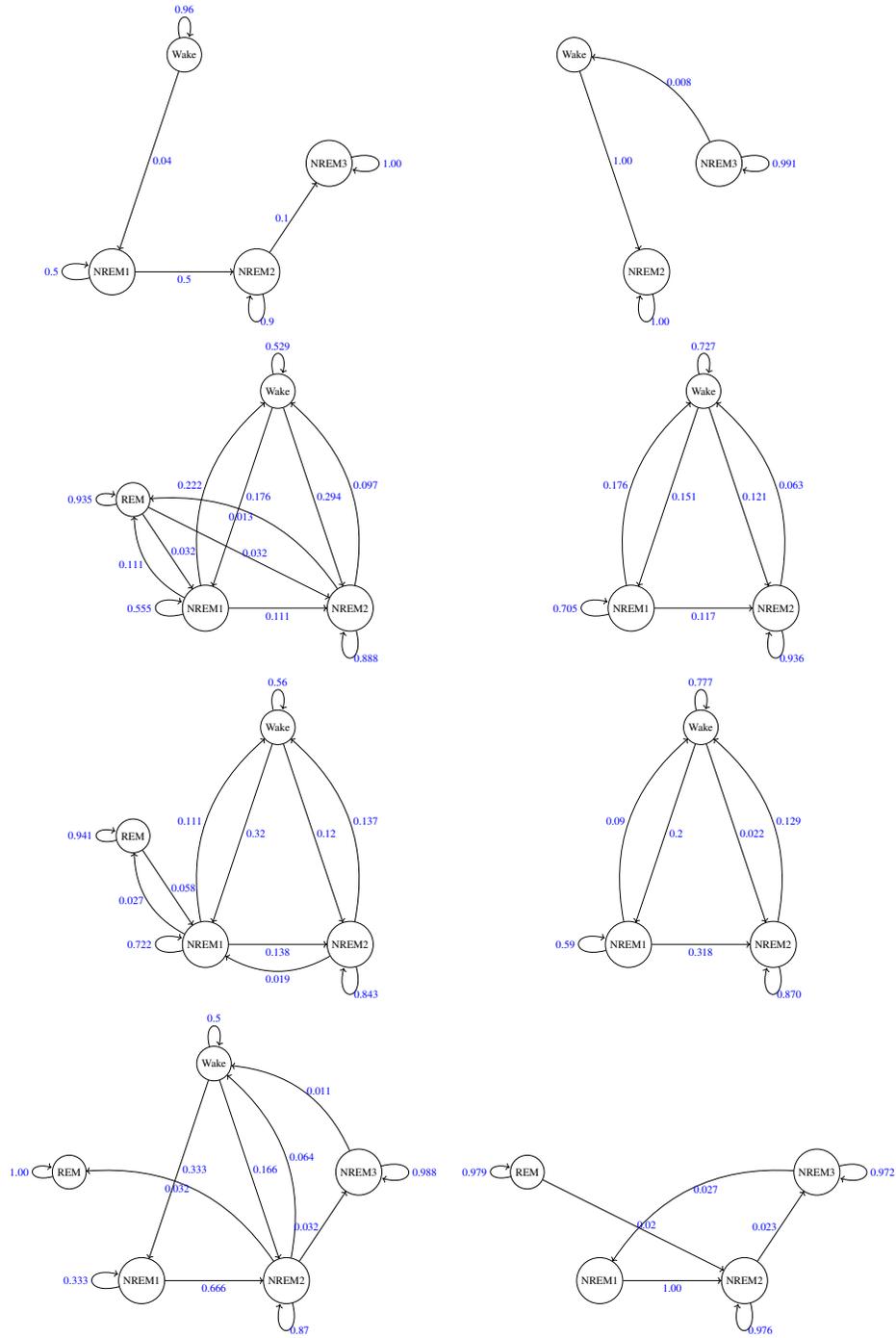

\begin{figure}
    \begin{tabular}{cc}
\scalebox{.5}{
\begin{tikzpicture}
\node[circle,draw,minimum size = 20pt] (a) at (0,0) {Wake};
\node[circle,draw] (c) at (-2,-6) {NREM1};
\node[circle,draw] (d) at (2,-6) {NREM2};
\node[circle,draw] (e) at (4,-3) {NREM3}; 
\path[->] (a) edge [loop above] node[blue] {0.866} (a);
\path[->] (a) edge node[right,blue]{0.133} (c);
\path[bend left,->] (c) edge node[left,blue]{0.076} (a);
\path[->] (c) edge [loop left] node[blue] {0.461} (c);
\path[->] (c) edge node[below,blue]{0.461} (d);
\path[bend right,->] (d) edge node[right,blue]{0.028} (a);
\path[bend left,->] (d) edge node[below,blue]{0.009} (c);
\path (d) edge [loop below] node[right,blue]{0.943} (d);
\path[->] (d) edge node[left,blue]{0.018} (e);
\path[bend right,->] (e) edge node[right,blue]{0.022} (a);
\path[bend left,->] (e) edge node[right,blue]{0.022} (d);
\path[->] (e) edge [loop right] node[right,blue]{0.955} (e);
\end{tikzpicture}
}
 & 
\scalebox{.5}{
\begin{tikzpicture}
\node[circle,draw,minimum size = 20pt] (a) at (0,0) {Wake};
\node[circle,draw] (c) at (-2,-6) {NREM1};
\node[circle,draw] (d) at (2,-6) {NREM2};
\path[->] (a) edge [loop above] node[blue] {0.903} (a);
\path[->] (a) edge node[right,blue]{0.096} (c);
\path[bend left,->] (c) edge node[left,blue]{0.058} (a);
\path[->] (c) edge [loop left] node[blue] {0.411} (c);
\path[->] (c) edge node[below,blue]{0.529} (d);
\path[bend right,->] (d) edge node[right,blue]{0.08} (a);
\path[bend left,->] (d) edge node[below,blue]{0.01} (c);
\path (d) edge [loop below] node[right,blue]{0.9} (d);
\end{tikzpicture}
}
 \\
\scalebox{.5}{
\begin{tikzpicture}
\node[circle,draw,minimum size = 20pt] (a) at (0,0) {Wake};
\node[circle,draw] (b) at (-4,-3) {REM};
\node[circle,draw] (c) at (-2,-6) {NREM1};
\node[circle,draw] (d) at (2,-6) {NREM2};
\node[circle,draw] (e) at (4,-3) {NREM3}; 
\path[->] (a) edge [loop above] node[blue] {0.5} (a);
\path[->] (a) edge node[right,blue]{0.5} (c);
\path (b) edge [loop left] node[blue] {0.986} (b);
\path[->] (b) edge node[right,blue]{0.013} (d);
\path[->] (c) edge node[below,blue]{1.00} (d);
\path[bend right,->] (d) edge node[right,blue]{0.013} (a);
\path[bend right,->] (d) edge node[left,blue]{0.013} (b);
\path (d) edge [loop below] node[right,blue]{0.945} (d);
\path[->] (d) edge node[left,blue]{0.027} (e);
\path[bend left,->] (e) edge node[right,blue]{0.016} (d);
\path[->] (e) edge [loop right] node[right,blue]{0.983} (e);
\end{tikzpicture}
}
 &
\scalebox{.5}{
\begin{tikzpicture}
\node[circle,draw,minimum size = 20pt] (a) at (0,0) {Wake};
\node[circle,draw] (b) at (-4,-3) {REM};
\node[circle,draw] (c) at (-2,-6) {NREM1};
\node[circle,draw] (d) at (2,-6) {NREM2};
\node[circle,draw] (e) at (4,-3) {NREM3}; 
\path[->] (a) edge [loop above] node[blue] {0.8} (a);
\path[->] (a) edge node[right,blue]{0.2} (c);
\path (b) edge [loop left] node[blue] {0.96} (b);
\path[->] (b) edge node[right,blue]{0.04} (d);
\path[bend left,->] (c) edge node[left,blue]{0.5} (a);
\path[->] (c) edge [loop left] node[blue] {0.5} (c);
\path[bend left,->] (d) edge node[below,blue]{0.04} (c);
\path (d) edge [loop below] node[right,blue]{0.96} (d);
\path[bend right,->] (e) edge node[above,blue]{0.006} (b);
\path[->] (e) edge [loop right] node[right,blue]{0.993} (e);
\end{tikzpicture}
}
    \end{tabular}
    \caption{Transition probabilities for the 4 sleep cycles of CF050 (displayed in order from left to right and top to bottom).} \label{fig:TP2}
\end{figure}
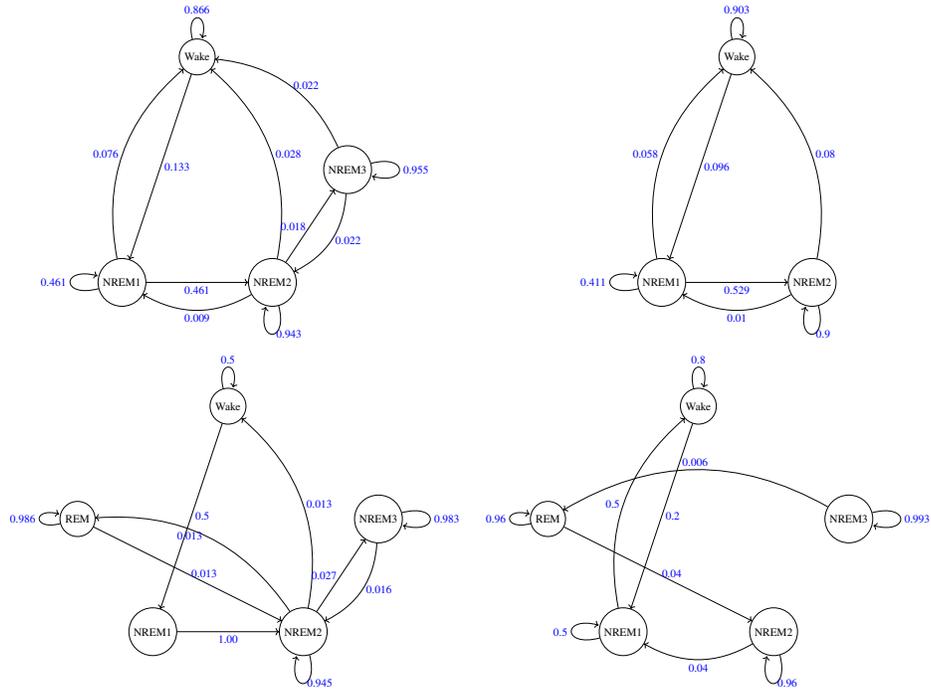

\clearpage
\bibliographystyle{plain}
\bibliography{ReferencesTime}
\end{document}